\documentclass[aps,preprint,graphicx,longbibliography,superscriptaddress]{revtex4-1}
\usepackage{dcolumn}
\usepackage{bm}
\usepackage{natbib}
\usepackage[usenames,dvipsnames] {color}
\usepackage[table]{xcolor}
\usepackage{graphicx}
\usepackage{epstopdf}
\usepackage{amssymb}
\usepackage{amsmath}
\usepackage{dcolumn}
\usepackage{subfigure} 
\DeclareGraphicsExtensions{.eps,.png,.pdf}
\usepackage{hyperref}
\hypersetup{
  colorlinks   = true, 
  urlcolor     = blue, 
  linkcolor    = blue, 
  citecolor   = blue 
}  
\draft 
\usepackage{setspace}
\usepackage[font=singlespacing,font=footnotesize, justification=RaggedRight]{caption}
\usepackage{enumitem}
\setlist{nolistsep}
\begin{document}

\title{Self-Propelled  Colloidal Particle Near a Planar Wall: A Brownian Dynamics Study} 
\author{Ali Mozaffari}
\affiliation{Benjamin Levich Institute and Department of Chemical Engineering, City College of the City University of New York, New York, NY 10031, USA.} \thanks{Current address: Institute for Molecular Engineering, University of Chicago, Chicago, Illinois 60637, USA.}
\author{Nima Sharifi-Mood}
\affiliation{Siemens PLM Software, Bellevue, WA 98005, USA.}
\author{Joel Koplik}
\email {jkoplik@ccny.cuny.edu.}
\affiliation{Benjamin Levich Institute and Department of Physics, City College of the City University of New York, NY 10031, USA.}
\author{Charles Maldarelli}
\email {cmaldarelli@ccny.cuny.edu.}
\affiliation{Benjamin Levich Institute and Department of Chemical Engineering, City College of the City University of New York, New York, NY 10031, USA.}

\date{\today}

\begin{abstract}
\begin{singlespace}
\footnotesize{Miniaturized, self-propelled locomotors use chemo-mechanical transduction mechanisms to convert fuel in the environment to autonomous motion.   Recent experimental and theoretical studies demonstrate that these autonomous engines  can passively follow the contours of solid boundaries they encounter.  Boundary guidance, however, is not necessarily stable: Mechanical disturbances can cause the motor to hydrodynamically depart from the passively guided pathway. Furthermore, given the scaled-down size of micromotors (typically 100 nm -10 $\mu$m), Brownian thermal fluctuation forces are necessarily important and these stochastic forces can randomize  passively-steered  trajectories. 

Here we examine theoretically the stability of boundary guided motion of micromotors along infinite planar walls to mechanical disturbances and to Brownian forces. Our aim  is to understand under what conditions this passively guided motion is stable. We choose a locomotor design in which spherical colloids are partially coated with a catalytic cap that reacts with solute to produce a product. The product is repelled from the particle surface,  causing the particle to move with the inert face at the front (autonomous motion via self-diffusiophoresis).   When propelled towards a planar wall, deterministic hydrodynamic studies demonstrate that  these locomotors can exhibit, for large enough cap sizes, steady  trajectories in which the particle either skims unidirectionally along the surface at a constant distance from the wall, or becomes stationary.  We first investigate the linear hydrodynamic stability of these states by expanding the equations of motion about the states, and find that linear perturbations decay exponentially in time. We then study the effects of thermal fluctuations by formulating a Langevin equation for the particle motion which includes the Brownian stochastic force. The Pe\'clet number scales the ratio of deterministic to Brownian forces, where ${\rm P}e = \pi \mu {a^2}{\tilde v_c}/{k_B}T$ and $a$ denotes the colloid radius, $\mu$ the continuous phase viscosity, ${\tilde v_c}$ the characteristic diffusiophoretic velocity and $k_{B}T$ the thermal energy. The skimming and stationary states are found to persist for ${\rm P}e$ above $10^{3}$.  At ${\rm P}e$ below $200$, the trajectory of a locomotor approaching the wall is unpredictable.  We present representative individual trajectories along with probability distributions for statistical ensembles of particles, quantifying the effects of thermal fluctuations and illustrating the transition from unpredictable to passively guided motion.}
\end{singlespace}
\end{abstract}

\maketitle

\section{Introduction}\label{intro}
The navigating  microbots of the future are currently being developed as synthetic,  self-propelled colloids, mirrors of motile micro-organisms,  which react with solute fuel in the environment, and convert the chemical energy of the reaction into mechanical motion. Technologies envision dispatching  these locomotors into small scale environments  to transverse the locales in applications  including the transport of cargo in microfluidic cells, the delivery of therapeutics to targeted sites in an organism,  tissue or even to the interior cytoplasm of a cell,  the scavenging of environs  for contaminants or cells, and the ferrying of macromolecules or nanoscale objects to a structure under assembly (for perspectives of the design of these engines and their applications see \cite{oz2005, Paxton:2006gf, Wang2009, WM2009,ms2009, SM2009, eh10, ozin2010, C0CS00078G, PUM2010, volpe2011,gibbszhao2011, sIS2012, sen2013, kap2013, marchetti2013hydrodynamics, buttinoni2013, wangreview2014,Schmidtr2014, Merkoci2014,degen2014,herminghaus2014interfacial, sen2015,sen2015a, Mallouk2015, sanchez2015, sancheza2015, Dey2016, ebbens2016active, bechinger2016,zottl2016, liu2016bimetallic, sen2017, Posner-rev, fei2017active}). In the current design of microbot locomotion, chemical energy is converted to mechanical motion by incorporating a catalyst on part of the surface of the colloid locomotor. The catalyst mediates the reaction of fuel in the environment to product species, and both become asymmetrically distributed around the colloid. The gradients in reactant and product solutes generate propulsion through interaction with the colloid, e.g. repulsive hard sphere or attractive van der Waals interactions for uncharged solutes (neutral diffusiophoresis), or electrostatic interactions in double layers of charge surrounding the motor for  ionic solutes and colloids with surface charge (ionic diffusiophoresis). The interaction generates a stress perpendicular to the colloid surface in the fluid layer around the locomotor \cite{anderson1982}.  Within this layer - typically thin relative to the colloid size - the  stresses  are balanced by pressure, and the resultant pressure gradients along the surface due to the asymmetric solute distributions of reactant and product drive a slip flow which propels the locomotor.  In addition, for colloids with surface charge and for ionic solutes, if the  solutes have different diffusivities or the colloid is conducting,  their asymmetric distribution can  create an electric field across the particle which  acts on the  particle double layer to electrophoretically propel the colloid (self-electrophoresis).  The products can also phase separate to form  bubbles which apply a propulsive thrust. Experimental and continuum theoretical studies of these propulsion mechanisms have been undertaken for locomotors in unbounded media, including for the mechanism of  self- neutral diffusiophoresis \cite{golestanian2005propulsion, 1367-2630-9-5-126, PhysRevLett.99.048102, Cordova, refId, Brady, golestanian_size, Ubaldo_Flux, campbell2013gravitaxis, sabass2012dynamics, Nima_1, Chemical_Sailing2014,frankel2014, takatori2014, yariv2015a, de2015diffusiophoretic,Takatori2016,Popescu2016, Popescu2017a,schmieding2017}, for  ionic diffusiophoresis \cite{yariv2010, seifert2012, poon2014, Golestanian2014, velegol2016origins, brown2017ionic, ibrahim2017multiple, wu2017zeta}, auto-electrophoresis \cite{PhysRevE.81.065302, posner2011, Posner2014, NourhaniBorhan2015, NourhaniBorhan2015a,Nourhanilammert2015, Posner-rev} and for bubble propulsion \cite{gibbs:163104, zhao2013, wang2014selecting}. See also the review by Balazs \emph{et al.} \cite{nikolov2016} for molecular dynamics perspectives on calculating the diffusiophoretic propulsion.

The ability to guide self-propelling microbots  through the landscape of a microenvironment is central to many of the above applications, and although  steering mechanisms based on using externally applied fields to manipulate the locomotor have been demonstrated (e.g. magnetic fields to maneuver magnetized locomotors), microbots designed to passively maneuver around objects or surfaces are of interest  since external fields do not have to be defined through the locale. Here, we study theoretically, using a continuum framework,  the passive guidance of locomotors along a planar wall, as a few key experimental studies - Sanchez \emph{et al.} \cite{simmchen2016}, Ebbens \emph{et al.} \cite{das2015} and Erbe \emph{et al.} \cite{erbe2013} - have observed this guidance experimentally as a steady skimming along a flat surface. These studies fabricated locomotors by coating one side of spherical  polystyrene colloids a few microns in diameter  with a platinum catalyst (a Janus locomotor), and fueled the motion by immersing the motor in a hydrogen peroxide solution. The catalyst decomposes the peroxide to oxygen and water; under high concentrations bubbles are produced which provide propulsion, but under the circumstances of these experiments, with the peroxide  concentrations relatively  low, bubble propulsion is secondary.  Proposed propulsion mechanisms include neutral self-diffusiophoresis of the interactions of the peroxide and oxygen with the colloid \cite{golestanian2005propulsion, 1367-2630-9-5-126, PhysRevLett.99.048102, refId, Brady, golestanian_size},  and electrokinetic mechanisms  (ionic self-diffusiophoresis and self-electrophoresis), driven by disassociation of the peroxide in the bulk or electrochemical reactions on the platinum surface \cite{seifert2012, poon2014, Golestanian2014, brown2017ionic, ibrahim2017multiple, wu2017zeta}.  Irrespective of the mechanism, the results are striking as Janus locomotors are observed to move along the wall, with their rotation suppressed and the normal to their active site approximately perpendicular to the normal to the wall. The movement is not rectilinear, but resembles 2D diffusion of the locomotor at an apparent constant distance from the wall \cite{isa2017}.

\begin{figure} [!htb]
\centering
\includegraphics[width=1.0 \textwidth]{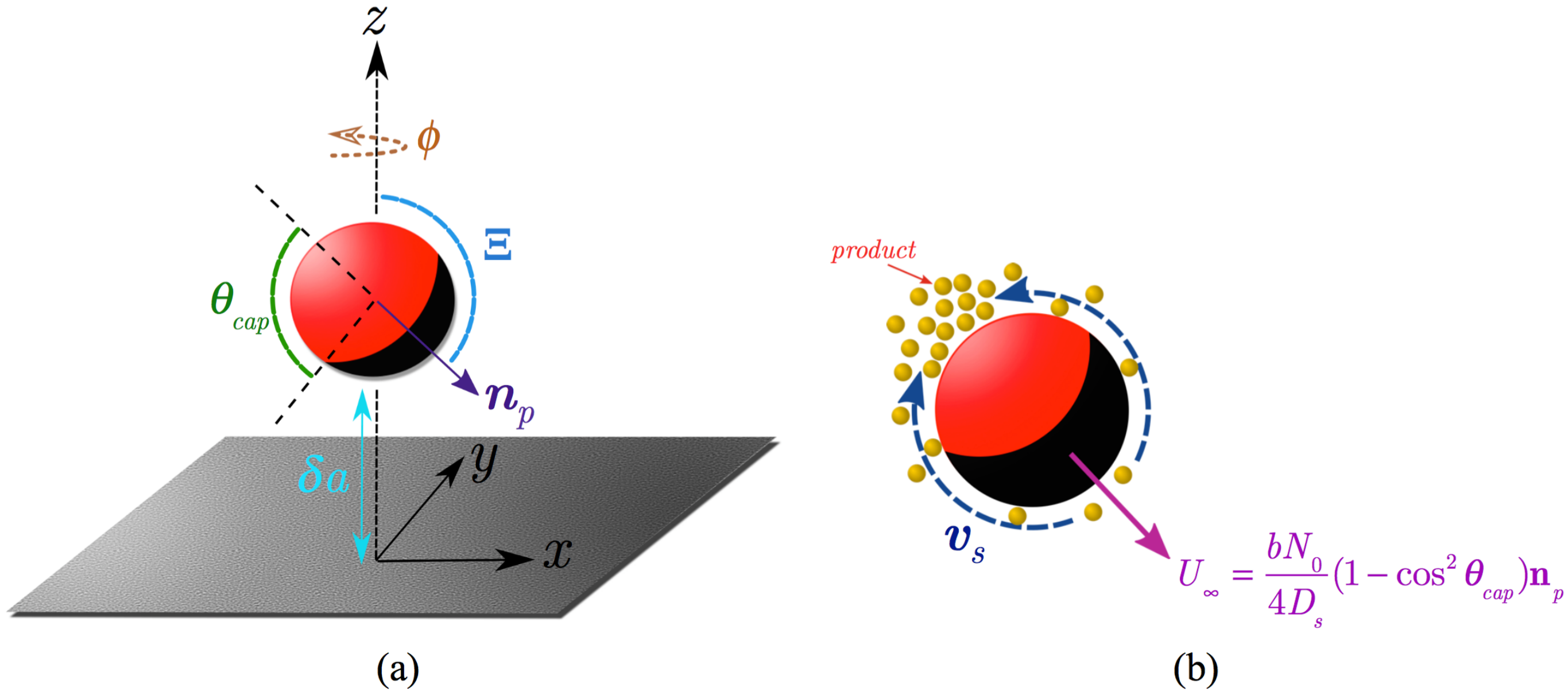}     	    
\caption{\footnotesize{(a) Schematic of  a Janus locomotor near a wall and  (b) locomotion in free space due to repulsive interaction of the product with the colloid.}} 
\label{schematic}
\end{figure}

Continuum theoretical studies to elucidate the  balance between the concentration-driven propelling forces and torques and the resistive hydrodynamic stresses and torques which can explain locomotor skimming were undertaken by Uspal \emph{et al.} \cite{Uspal} using boundary integral numerical solutions, Ibrahaim and Liverpool \cite{liverpool2015, ibrahim2016walls} (using singularity methods) and Mozaffari \emph{et al.} \cite{mozaffari2016} using analytical  solutions in bispherical coordinates.  All these authors modeled, as in the experiments,  spherical Janus locomotors with a symmetrical reactive cap (angle $\theta_{cap}$) (Fig. \ref{schematic}a),  and assumed that the propulsion was driven solely by neutral self-diffusiophoresis (see also Crowdy's analysis of two dimensional Janus circles \cite{crowdy2013}, Qiao \cite{yang2016} for motion of a Janus particle in a cylindrical pore and \cite{sharifi2016, papavassiliou2017} for two interacting swimmers). For simplicity, only the interaction of one species of the reaction scheme (e.g. the product)  generated the propulsion, interacting with the colloid with an intermolecular potential $\phi(r)$, where $r$ is the separation distance between the solute and the colloid. The effective range of the interaction was assumed to be of order $L$, much smaller than the radius of the particle $a$, and therefore the propulsion was described by a tangential slip velocity ($\bm{\tilde{v}}_{s}$) at the surface of the colloid as given by $\displaystyle{\bm{\tilde{v}}_{s}  =  - b\bm{\tilde{\nabla}} _s \tilde{C}}$, where $\tilde C$ is the concentration field outside the particle, $\tilde{ \nabla}_{s}$ is the surface gradient and $b$ is the slip coefficient ($\displaystyle{b = \frac{{{k_B}T{L^2}}}{\mu }\int_0^\infty  {y\left[ {\exp \left( { - \phi (y)/{k_B}T} \right) - 1} \right]dy}}$) in which $\mu$ is the viscosity of the continuous phase, $k_{B}T$ is the thermal energy and $y=(r-a)/L$ is the rescaled distance from the colloid. (A ``tilde'' indicates a dimensional quanitity.) Repulsive interactions were considered ($b<0$) for which the slip velocity is in the direction of the concentration gradient.  The reaction rate was assumed to be slow relative to diffusion so that the flux of product from the surface  was constant ($N_{o}$) \cite{yariv2015,Lauga_Peclet}, and the product did not penetrate or interact with the wall, 
and the flow produced by the slip velocity was assumed to be at low Reynolds number (zero inertia) and Pe\'clet number (convection of solute negligible) due to the small size of the locomotors. In the absence of Brownian forces, due to the symmetry of the colloid, the locomotor  translates in a plane formed by its symmetry axis (here taken to be the unit normal $\bm {n}_{p}$ from the center of the inactive area) and the unit normal to the wall which is coincident with the $z$ axis. Far from the surface, the locomotor translates rectilinearly in the direction $\bm{n}_{p}$ with velocity $\tilde U_{\infty}=\displaystyle{{\tilde v_c}}(1 - \cos^2~{\theta _{cap}})/4$ where $\displaystyle{\tilde v_{c}=b{N_{o}}/{D_{s}}}$ and $D_{s}$ is the diffusion coefficient of the product (Fig. \ref{schematic}b).  For a given distance between the locomotor and the wall ($\delta a$) and  inclination  of the cap of the locomotor with the wall ($\Xi$, $\cos \Xi = \bm {n}_{p} \cdot \bm{\hat e}_z$ ), master curves of the (nondimensional) translational velocities in the $z$ and $x$ directions ($U_{\perp}(\delta, \Xi;\theta_{cap})$ and $U_{\parallel}(\delta, \Xi;\theta_{cap})$, both scaled  by $\tilde v_{c}$)  and nondimensional rotation rate about the $y$ axis  ($\Omega(\delta, \Xi;\theta_{cap})$, scaled by $\tilde v_{c}/a$) were obtained.  Using these master curves, simulations of locomotors approaching a wall (inclination angles $\Xi$ larger than $90^{o}$) demonstrated that for a  range of large active areas ($115^{o} < \theta_{cap} < 145^{o}$) the locomotors rotated to an inclination angle ($\Xi_s(\theta_{cap})$) and moved to a gap thickness ($\delta_{s}(\theta_{cap})$) for which rotation was suppressed ($\Omega(\delta_{s}, \Xi_{s};\theta_{cap})=0$) and the locomotor skimmed along the surface ($U_{\perp}(\delta_{s}, \Xi_{s};\theta_{cap})=0$), translating only in the $x$ direction.
\begin{figure} [!htb]
\centering
\includegraphics[width=1.0 \textwidth]{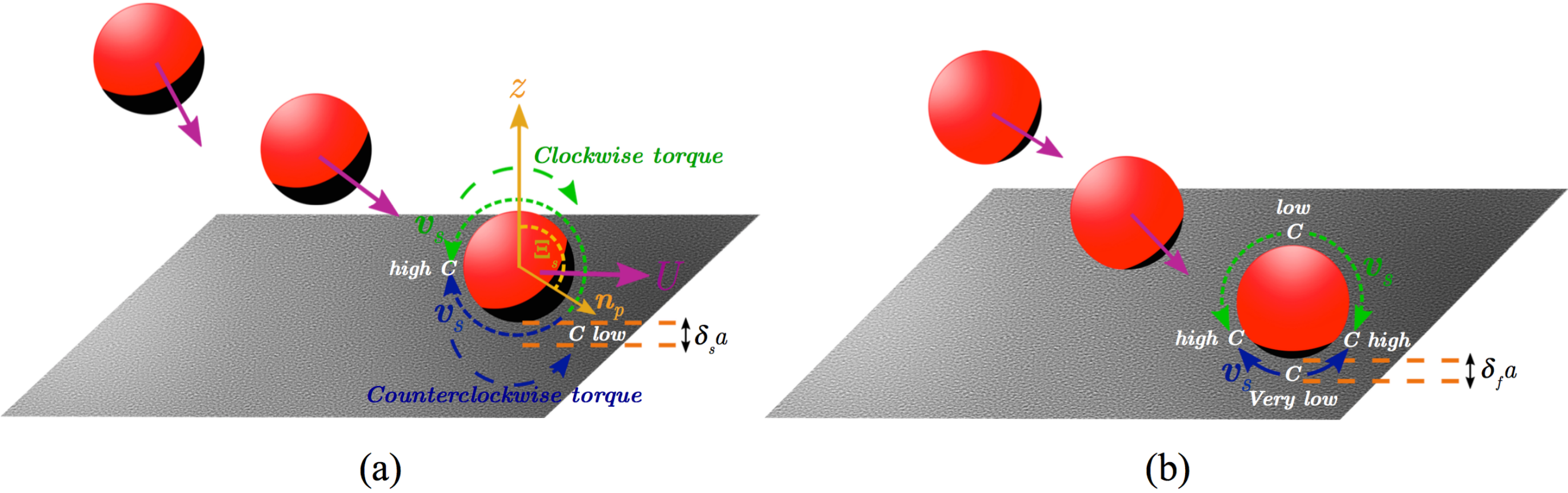}     	    
\caption{\footnotesize{(a) Skimming state of Janus locomotor along a surface for large active caps and  (b) Stationary state of Janus locomotor for very large active caps.}} 
\label{schematic2}
\end{figure}

The concentration field of the product around the locomotor is shown in Fig. \ref{schematic2}a (see Mozaffari \emph{et al.} \cite{mozaffari2016}); the large active area results in build-up of product in the vicinity of the gap between the motor and the wall, which leads to a maximum in the concentration at the locomotor surface which is close to the wall and away from the center of the active area.  The slip velocities and resultant propulsive  shear stresses applied to the locomotor are therefore asymmetrically distributed around the colloid. These stresses result in motion along the wall ($x$ axis). The stresses also result in propulsive torques on the colloid about the $y$ axis. In particular, the slip on the face of the colloid nearest the wall (solid delimiter) acts to rotate the locomotor  in the counterclockwise direction, and the slip on the opposing face (dotted line delimiter) acts to rotate in the clockwise direction. Recall that forward motion of the colloid also generates a hydrodynamic induced clockwise torque. In the skimming state, the counterclockwise propulsive torque applied by the lower face is larger than the clockwise torque applied by the upper face due to the  proximity of the lower face to the wall which generates large shear forces. This net propulsive counterclockwise action balances the clockwise hydrodynamic torque due to the forward rotation, and net rotation is suppressed.   The shear stresses from the slip velocities also  generate propulsive forces in the $z$ direction, and these are largest on the side of the colloid where  the maximum in concentration is located due to the fact that the maximum is inclined to the wall. In the skimming state,  the shear from the slip below the maximum creates a propulsive force in the negative $z$ direction which balances the propulsive force in the positive $z$ direction due to the shear from the slip above the maximum. The interaction of the concentration field on suppressing the rotation is unique to chemically-powered locomotors; however colloidal models with prescribed slip conditions on their surfaces (i.e. ``squirmers'' or pushers or pullers which are used to describe the mechanics of motile microorganisms) have also demonstrated skimming for suitable slip velocities (see Spagnolie and Lauga \cite{Lauga_Spagnolie} and Ishimto and Gaffney \cite{ishimoto2013} and the reviews \cite{lauga2009a,elgeti2016}).

The numerical studies \cite{Uspal,mozaffari2016,liverpool2015, ibrahim2016walls} also showed that for even larger active areas than those which give rise to skimming, i.e. $145^{\circ}<\theta_{cap} < 180^{\circ}$, a locomotor,  again approaching a wall, rotates clockwise to a symmetric position ($\Xi =180^{o}$) and gap distance ($\delta_{f}$) for which the locomotor stops (a stationary state). The concentration field in the stationary state is shown in Fig. \ref{schematic2}b. The large active area results in a large accumulation in the gap between the locomotor and the wall, and a maximum in concentration along the locomotor surface which extends symmetrically  as a belt around the locomotor and close to the wall. The high concentration perimeter in the stationary state is closer to the wall than the maximum in  the skimming state. The slip velocites are directed from the locomotor pole facing away from the wall to the belt, and the pole facing the wall to the belt. The corresponding shear stresses propel the particle upward (due to the slip on the top half) and downward (slip on the bottom half). Although the slip on the upper half occupies a larger area, the slip on the lower half creates a greater shear due to the proximity to the wall and results, at a critical gap thickness, in a propulsion downward that balances the upward propulsion of the top half so that the particle becomes stationary.  
 
Additional theoretical studies on the guidance at planar walls of Janus locomotors propelled by neutral self-diffusiophoresis have investigated the effect of shear flow on the skimming and stationary states \cite{Tasinkevych2015}, directed guidance along walls by the modification of the surface with a gradient in surface chemistry such that the reactant and product solutes interact with the wall to create an additional chemo-osmotic flow \cite{popescu2017}, guidance along steps on a flat surface \cite{uspal2016guiding} and the induction of propulsion of Janus locomotors with homogeneous coatings of catalyst due to the distortion of the solute concentration fields around the particle by an inert planar (nonadsorbing) surface \cite{michelin2015,yariv2017,yariv2016}. 

All the above calculations for the skimming and stationary states of the diffusiophoretic Janus locomotors were undertaken in the deterministic limit by solving the equations for the mass conservation of the solutes and the equations of motion for fluid (in the Stokes flow regime) and the particle itself. In any realistic situation, other particles, contaminants or even mechanical perturbations will be present and the hydrodynamic stability of the skimming and stationary trajectories is an issue.  Furthermore, and more generally, colloidal particles exhibit Brownian motion due to thermal fluctuations in the fluid environment which may obliterate deterministic motion  \cite{romanczuk2012,zottl2016,bechinger2016}. Intuitively, one expects Brownian effects to dominate for small or slow particles or high temperature, but more quantitative results are needed to utilize the self-propellers in the applications described above.  Brownian dynamics of simple active Brownian particles subject to a constant propulsion force or prescribed slip velocities (squirmers) and without hydrodynamic interactions  has been studied \cite{ten2011,redner2013,takatori2014,romanczuk2012,zottl2016} for isotropic and anisotropic particles and also for the collection of two dimensional micro-rods near a V shaped solid wall \cite{kaiser2012}. Theoretical and experimental analysis of Brownian motion of chemically powered Janus swimmer \cite{howse2007} revealed the transition from initial directed motion to diffusive regime at longer time. The effect of Brownian motion on the trapping time of self propelled bi-segmented Au-Pt rod  and also squirmers (pushers and pullers) which orbits around passive spherical particle and the comparison of swimming through colloidal crystal for E. coli bacteria and catalytic Janus swimmer has been studied \cite{takagi2014,spagnolie2015,brown2016}. The effect of thermal fluctuations on the distribution of detention time of active Brownian particles and also squirmers near a planar no-slip wall \cite{schaar2015,ruhle2017gravity} was theoretically studied. 

In this paper we examine the stability issues for skimming and stationary self diffusiophoretic particle motion along a wall.  The calculations closely follow the deterministic analysis in our previous paper \cite{mozaffari2016}, where the instantaneous solutions of the governing equations give the linear and angular velocity at each particle position and orientation, which are used to update the configuration by a time-stepping procedure.  We first consider the hydrodynamic stability of solutions to the continuum equations, and the result is that they are indeed linearly stable.  We then focus on the stability of the skimming and stationary states to thermal fluctuations. If we define a characteristic velocity for the diffusiophoretic propulsion of a colloid with radius $a$ by the value in an unbounded fluid,  $\tilde {v}_{c}$, then the characteristic diffusiophoretic force is this reference velocity multiplied by the viscous drag ($6\pi \mu a \tilde{v}_{c}$) where $\mu$ is the continuous phase viscosity. The energy expended by the diffusiophoretic force in moving the particle over its radius is  $6\pi \mu a^{2} \tilde{v}_{c}$. This energy is then compared to the thermal energy $k_{B}T$ to define a P\'eclet number ${\rm P}e={\pi \mu a^{2} \tilde{v}_{c}}/{k_{B}T}$ which therefore scales the deterministic propulsive energy to the thermal fluctuations. (Alternatively, ${\rm P}e= a\tilde{v}_{c}/(6\mathcal{D})$ where $\mathcal{D}$ is the colloid diffusion coefficient given by the Stokes-Einstein equation $\mathcal{D}=k_{B}T/6\pi \mu a$.) We anticipate that if the P\'eclet number is sufficiently large, then the deterministic states will be stable to Brownian motion.  Typical values of the P\'eclet number for colloidal propulsion can be obtained using  the hydrogen peroxide fueled catalytic motors as an example, for which $a$ in the range of 1-20 microns and $\tilde{v}_{c}$ is of order 1-10 microns/sec, which results in P\'eclet numbers in the range of one to order one thousand.  Hence we will use this range in our simulations. Colloid trajectories are obtained by using an overdamped Langevin equation which incorporates the propulsive diffusiophoretic force, hydrodynamic drag, and a stochastic force to account for the Brownian fluctuations \cite{ermak1978}. We compare  the deterministic trajectories to the ones obtained including the fluctuations forces as a function of the P\'eclet number to understand the stability of the  skimming and stationary states.

\section{Formulation and Analytical Solution}
\label{formulation} 
\subsection{Propulsive Diffusiophoretic Force and Torque}
\label{propulsiveformulation}
The force driving the motion of a colloid is the interaction with the solute concentration gradient around the surface, so the first step is to 
solve the mass transfer equation for the solute. In the limit of small solute P\'eclet number, the advection of solute can be neglected and if the 
reaction rate is negligible compared to the solute diffusion rate (small Damk\"ohler number) \cite{yariv2015,Lauga_Peclet,khair2013,natale2017} we may assume a uniform solute production rate ($N_0$) at the active face of colloid (and zero flux at the passive face). The governing equation and boundary conditions can then be written in dimensionless form, as: 
\begin{eqnarray}
  \nabla ^2  C = 0, \\ 
 {\bf{n}}\cdot \bm{\nabla} C = 0,\quad \quad {\bf{r}} \in \Gamma _W,  \\ 
  - {\bf{n}}\cdot \bm{\nabla} C = \left\{ \begin{array}{l}
 1\quad \quad \,{\bf{r}} \in \Gamma _{AP}  \\ 
 0\quad \quad {\bf{r}} \in \Gamma _{PP}  \\ 
 \end{array} \right., \\ 
 C \to 0,\quad \left| {\bf{r}} \right| \to \infty,  
\end{eqnarray}
Here $\Gamma_W$, $\Gamma_{AP}$ and $\Gamma_{PP}$ represent the surfaces of the wall, the active side of particle and the passive part of particle, 
respectively. The solute concentration $C(\bf{r})$ has been non-dimensionalized as $ C = \tilde C\,(N_0 a/D_s )^{-1}$, where $a$ is particle radius 
and $D_s$ is the solute diffusivity, and $\bm{\nabla} = a\bm{\tilde\nabla}$. The active cap geometry is specified by the coverage angle $\theta_{cap}$ and the tilt angle $\Xi$ and the gap distance $\delta$ between the particle and the wall, nondimensionalized by $a$ (Fig. \ref{schematic}a).  We solve for the solute concentration using the three-dimensional solution of Laplace's equation in bispherical coordinate.

The concentration couples to the fluid flow through the diffusiophoretic slip velocity boundary condition on the particle surface, $\bm{\tilde v}_s =-b\bm{\tilde\nabla} _s \tilde C$ were $b$ is the slip coefficient as defined in the Introduction (Sec.~\ref{intro}).   At low Reynolds number the Stokes and continuity equations apply, and their dimensionless form are
\begin{eqnarray}
 \nabla ^2 {\bf{v}} - \bm{\nabla} p = 0, \\ 
 \bm{\nabla} \cdot {\bf{v}} = 0, 
\end{eqnarray}
subject to the following boundary conditions
\begin{eqnarray}
 {\bf{v}} = - \bm{\nabla} _s  C,\quad \quad {\bf{r}} \in \Gamma _P,  \\ 
 {\bf{v}} = 0,\quad \quad {\bf{r}} \in \Gamma _W,  \\ 
 {\bf{v}} = 0,\,\,\,\,\quad \left| {\bf{r}} \right| \to \infty,  
\end{eqnarray}
where ${\bf{v} = \tilde v}/\tilde v_c$, $p = \tilde p\,(\mu \tilde v_c/a)^{-1}$ and the characteristic velocity for an unbounded locomotor is given (see Introduction, Sec.~\ref{intro}) by $\tilde v_c = b N_0/D_s$.  The traction of the flow field at the particle surface produces the propulsive diffusiophoretic force and torque via
\begin{eqnarray}
 {\bf{\tilde F}}_P  = \mathop{{\int\!\!\!\!\!\int}\mkern-21mu \bigcirc}\limits_{\Gamma _P } 
 {\left( {\bm{\tilde \sigma} \cdot {\bf{n}}} \right)} \,ds,
 \label{propulsiveforcetorque1} 
 \end{eqnarray}
 \begin{eqnarray}
 {\bf{\tilde T}}_P  = \mathop{{\int\!\!\!\!\!\int}\mkern-21mu \bigcirc}\limits_{\Gamma _P } 
 {\left( {{\bf{r}}_p  - {\bf{r}}_o } \right) \times \left( {\bm{\tilde \sigma} \cdot {\bf{n}}} \right)} \,ds.
 \label{propulsiveforcetorque2} 
\end{eqnarray}
Here, $\bm{\tilde \sigma}  =  - \tilde p{\bf{I}} + \mu \left[ {\bm{\nabla} {\bf{\tilde v}} + \left( {\bm{\nabla} {\bf{\tilde v}}} \right)^T } \right]$ is stress tensor, ${\bf{n}}$ is the outward unit vector normal to the particle surface $\Gamma _P$, and ${\bf{r}}_{p}$ and ${\bf{r}}_o$ are position vectors on the particle surface and at the particle's center of mass, respectively. For a given cap geometry,  the magnitudes of the diffusiophoretic force and torque are functions of the tilt angle and gap distance to the wall, and their orientations are as follows. If the unit vector $\bm{n}_{p}$ specifies the orientation of the particle, {\sl e.g}., a vector pointing from the center of the particle to the center of the inert face of the locomotor (Fig. \ref{schematic}a), and $\bm{\hat e}_z$ is the unit normal vector of the wall, then the force is in the plane formed by these two vectors and the torque is in the direction of $\bm{n}_{p} \times \bm{\hat e}_z$ . In Cartesian coordinates where the wall is the $x$-$y$ plane, $ {\bf{\tilde T}}_{P}\cdot {{\bm{\hat e}}_z}=0$ at all times.  The  solution for the propulsive force and torque are obtained by solution in bispherical coordinates and the details are in Mozaffari \emph{et al.} \cite{mozaffari2016}.

\subsection{Hydrodynamic Force and Torque}
\label{resistiveformulation}
The motion of a sphere moving near a wall with translational velocity $\bm{\tilde U}$ and rotational velocity $\bm{\tilde \Omega}$, can be decomposed into four independent simplified sub-problems (Fig.~\ref{resist}) due to the linearity of the Stokes equations. These are:
\begin{itemize}
\item
The sphere translates without rotation in a direction perpendicular to the wall (Fig.~\ref{resist}(a)). This translation results in a drag on the sphere in the direction perpendicular to the wall $f^{tt}_\perp$.
\item
The sphere translates without rotation in a direction parallel to the wall (Fig.~\ref{resist}(b)). This translation results in a drag parallel to the wall ($f^{tt}_\parallel$) and a torque about an axis parallel to the wall and perpendicular to the direction of motion ($f^{tr}_\parallel$).
\item
The sphere rotates without translation around an axis perpendicular to the wall (Fig.~\ref{resist}(c)). This rotation results in a torque around the axis perpendicular to the wall ($f^{rr}_\perp$).
\item
The sphere rotates without translation around an axis parallel to the wall (Fig.~\ref{resist}(d)). This rotation results in a drag force in a direction along the wall and perpendicular to the axis of rotation ($f^{rt}_\parallel$), and a torque around the axis of rotation ($f^{rr}_\parallel$).
\begin{figure} [!htb]
\centering
\includegraphics[width=0.90 \textwidth]{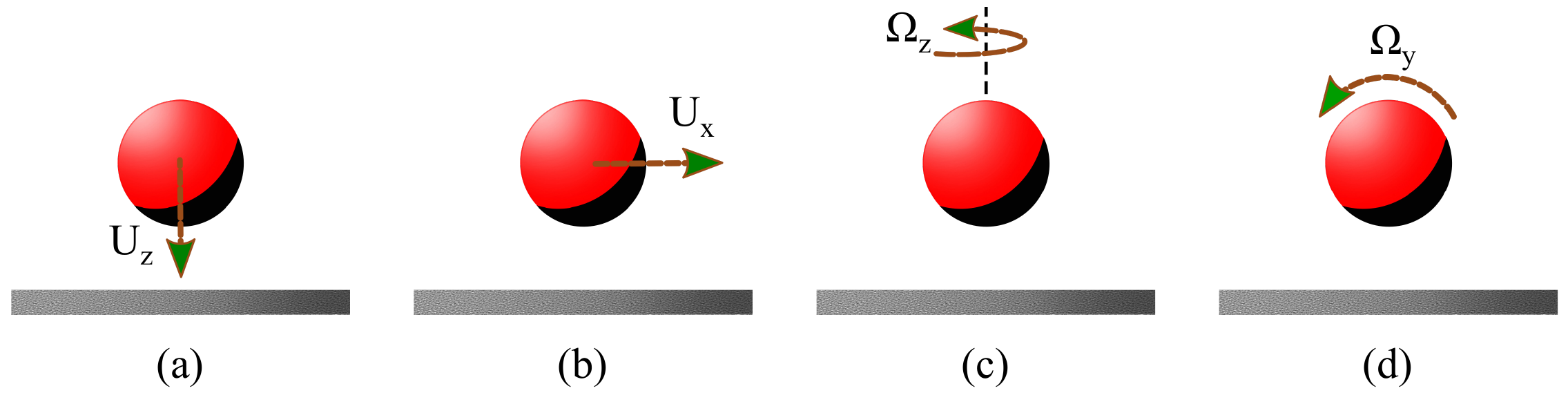}     	    
\caption{\footnotesize{Different modes of motion to construct the resistance matrix (resistance problems).}} 
\label{resist}
\end{figure}
\end{itemize} 
Thus the hydrodynamic resistance force ${\bf{\tilde F}}_H$ and torque ${\bf{\tilde T}}_H$  due to the translational and rotational motion of the colloid can be written in terms of the resistance matrix $\bm{\mathcal{\tilde R}} $ in a more concise form as: 
\begin{equation}
\left( \begin{array}{l}
{{{\bf{\tilde F}}}_H}\\
{{{\bf{\tilde T}}}_H}
\end{array} \right) =  - \bm{\mathcal{\tilde R}} \left( \begin{array}{l}
{{\bf{\tilde U}}}\\
{{\bf{\tilde \Omega }}}
\end{array} \right).
\label{hydrodynamicresistance}
\end{equation}
where the resistance matrix $\bm{\mathcal{\tilde R}}$ is defined in terms of the coefficients defined above by \cite{kim2013}:
\begin{equation}
\bm{\mathcal{\tilde R}}  = \pi \mu a\left( {\begin{array}{*{20}c}
   {6 f^{tt} _\parallel  } & 0 & 0 & 0 & { - 6 a f^{rt} _\parallel  } & 0  \\
   0 & {6 f^{tt} _\parallel  } & 0 & {6 a f^{rt} _\parallel  } & 0 & 0  \\
   0 & 0 & {6 f^{tt} _ \bot  } & 0 & 0 & 0  \\
   0 & {\quad 8 a f^{tr} _\parallel  } & 0 & {8 a^2 f^{rr} _\parallel  } & 0 & 0  \\
   {\quad  - 8 a f^{tr} _\parallel  } & 0 & 0 & 0 & {8 a^2 f^{rr} _\parallel  } & 0  \\
   0 & 0 & 0 & 0 & 0 & {8 a^2 f^{rr} _ \bot  } \\
\end{array}} \right),
\label{Rmatrix}
\end{equation}
The details of solution for each resistance problem can be found in our previous paper \cite{mozaffari2016}, except for the problem correspond to Fig.~\ref{resist}(c) which is discussed in
the Appendix~\ref{append1}. 
\subsection{Linear Stability Analysis}
To examine the linear hydrodynamic stability of the skimming (Fig. \ref{schematic2}a) and stationary (Fig. \ref{schematic2}b) states, we first formulate equations for the translational trajectory and rotational movement of the locomotor when it is disturbed from these base states by small  amplitude perturbations. With reference to the coordinate system of Fig. \ref{schematic}a, the locomotor's position and orientation are specified for a general position denoted by ``p'' by the location of its center $(x_{p}, y_{p}, z_{p})$, its inclination angle $\Xi_{p}$ from the $z$ axis, and the azimuthal angle $\phi_{p}$ around the $z$ axis.  Only five coordinates are necessary because the locomotor has an axis of symmetry, the axis passing through the center of the passive area and the locomotor center. This axis points in the direction $\bm{n}_{p}$ as shown in  Fig.~\ref{schematic}a. In addition, as we noted earlier, for a given  orientation $\Xi_{p},\phi_{p}$, the locomotor rotates around an axis (unit direction $\bm{\xi}_{p}$) which is  perpendicular to the $z$ axis and the body-fixed normal vector $\bm{n}_{p}$. 
\begin{align}
\bm{n}_p &= \sin \Xi_{p} \cos \phi_{p} \bm{\hat{e}}_x + \sin \Xi_{p} \sin \phi_{p} \bm{\hat{e}}_y + \cos \Xi_{p} \bm{\hat{e}}_z\\
\bm{\xi} _p & =  \cos \phi_{p} \bm{\hat{e}}_y - \sin \phi_{p} \bm{\hat{e}}_x
\end{align}
At this position ``p''  the locomotor is translating with a velocity with components normal to the wall ($U_{\perp}(\delta_{p},\Xi_{p})$) in direction $\bm{\hat{e}}_z$ and parallel to the wall ($U_{\parallel}(\delta_{p},\Xi_{p})$) in the direction $\bm{\xi}_{p} \times \bm{\hat{e}}_z$, and rotating about $\bm {\xi}_{p}$ with rotational velocity $\Omega(\delta_{p},\Xi_{p})$ where $\delta_{p}$ is the gap distance from the wall, $z_{p}-1$. 
\begin{align}
\bm{U} &  =  U_{\|}(\delta_{p},\Xi_{p}) \cos \phi_{p} \bm{\hat{e}}_x + U_{\|}(\delta_{p},\Xi_{p})  \sin \phi_{p} \bm{\hat{e}}_y + U_{\perp}(\delta_{p},\Xi_{p}) \bm{\hat{e}}_z \\
\bm{\Omega} & = \Omega(\delta_{p},\Xi_{p}) (\cos \phi_{p} \bm{\hat{e}}_y - \sin \phi_{p} \bm{\hat{e}}_x)
\end{align}
The trajectory of the particle at position ``p'' can be constructed by integrating the  kinematic equations:
\begin{align}
&\frac{{\;d{x_p}}}{{dt}}  = U_{\|}(\delta_{p},\Xi_{p}) \cos \phi_{p},\label{translationeqs1}\\ 
&\frac{{\;d{y_p}}}{{dt}} = U_{\|}(\delta_{p},\Xi_{p}) \sin \phi_{p},\\
&\frac{{\;d{z_p}}}{{dt}}  = U_{\perp}(\delta_{p},\Xi_{p}) ).\label{translationeqs}
\end{align}
The  rotational  kinematics can be obtained by integrating:
\begin{eqnarray}
\frac{d\bm{n}_p}{dt} = \bm{\Omega} \times \bm{n}_p,
\end{eqnarray}
which leads to
\begin{align}
\frac{{d{\phi _p}}}{{dt}} &= 0,\\
\frac{{d{\Xi _p}}}{{dt}} &= {\Omega}(\delta_{p},\Xi_{p}).
\label{rotationeqs}
\end{align}
Consider first the linear stability of the skimming state. For a particular range of cap angles ($115^\circ < \theta_{cap} < 145^\circ$), the locomotor skims in rectilinear motion parallel to the wall  with its active area at a fixed  inclination $\Xi_{s}$ and separation distance $\delta_{s}$, where each of these are a function of $\theta_{cap}$.  Small deviations from this  trajectory can be described by deviation functions denoted by $\delta$:
\begin{eqnarray}
{x_{(skim)}}(t)& = \cos {\phi _s}U_{\parallel}({\delta _s},{\Xi _s})~t + \delta {x_{(skim)}},\\
{y_{(skim)}}(t) &= \sin {\phi _s}U_{\parallel}({\delta _s},{\Xi _s}) ~t + \delta {y_{(skim)}},\\
{z_{(skim)}}(t) &= 1 + {\delta _s} + \delta {z_{(skim)}},\\
{\Xi _{(skim)}}(t) &= {\Xi _s} + \delta {\Xi _{(skim)}},\\
{\phi _{(skim)}}(t)& = {\phi _s} + \delta {\phi _{(skim)}},
\end{eqnarray}
where we assume that the skimming direction in the $x-y$ plane is described by $\phi_{s}$, which is arbitrary and at $t=0$ the center of the locomotor is above the origin of the cartesian system $(x=0, y=0, z=1+\delta_{s})$. The deviation functions have initial values that represent the perturbation imposed on the skimming state. Substituting the above definitions of the deviation functions  into the kinematic equations \ref{translationeqs1}$-$\ref{rotationeqs} yields a set of dynamical system relations  for the deviation functions after linearization:
\begin{equation}
\begin{split}
\frac{d}{{dt}} \left[\delta x_{skim} \right] =   &\cos \phi _{s}  {\left[ {\frac{{\partial U_{\|}}}{{\partial \delta }}} \right]_{|{\delta _{s}},\Xi _{s}}} \left[\delta z_{skim} \right ]
                                                +\cos \phi _{s}  {\left[ {\frac{{\partial U_{\|}}}{{\partial {\Xi _p}}}} \right]_{|{\delta_{s}},\Xi _{s}}} \left[ \delta \Xi_{skim} \right]\\
                                                &- \sin \phi _{s}U_{\|}(\delta _{s},\Xi _{s}) \delta \phi_{skim}(t=0),
\end{split}
\label{xskim}
\end{equation}
\begin{eqnarray}
\begin{split}
\frac{d}{{dt}}\left[\delta y_{skim} \right] =   &{\sin \phi _{s}\left[  {\frac{{\partial U_{\|}}}{{\partial \delta }}} \right]_{|{\delta _{s}},\Xi _{s}}} \left[  \delta z_{skim} \right]
                                                + \sin \phi _{s}{\left[ {\frac{{\partial U_{\|}}}{{\partial {\Xi _p}}}} \right]_{|{\delta_{s}},\Xi _{s}}} \left[  \delta \Xi_{skim} \right] \\ 
                                                 &+ \cos \phi _{s} U_{\|}(\delta _{s},\Xi _{s})\delta \phi_{skim}(t=0),
\end{split}  
\label{yskim}                                             
\end{eqnarray}
\begin{eqnarray}
\begin{split}
\frac{d}{{dt}} \left[ \delta z_{skim} \right] =  {\left[ {\frac{{\partial U_{\perp}}}{{\partial \delta }}} \right]_{|{\delta _{s}},\Xi _{s}}} \left[ \delta z_{skim} \right]
                                               + {\left[ {\frac{{\partial U_{\perp}}}{{\partial {\Xi _p}}}} \right]_{|{\delta _{s}},\Xi _{s}}} \left[ \delta \Xi_{skim}  \right ],     
 \end{split}          
 \label{zskim}
 \end{eqnarray}
 \begin{eqnarray}
 \begin{split}
\frac{d}{{dt}} \left[ \delta \Xi_{skim} \right ]= {\left[ {\frac{{\partial {\Omega}}}{{\partial \delta }}} \right]_{|{\delta _{s}},\Xi _{s}}}\left[\delta z_{skim} \right] 
 + {\left[ {\frac{{\partial {\Omega }}}{{\partial {\Xi _p}}}} \right]_{|{\delta _{s}},\Xi _{s}}}\left[\delta \Xi_{skim}\right].
\end{split}
\label{xiskim}
\end{eqnarray}

The solutions for $\delta y_{skim}$  and $\delta \Xi_{skim}$ can be determined directly from Eqs.~\ref{zskim} and \ref{xiskim} in the form: 
\[\left[ \begin{array}{l}
\delta {z_{(skim)}}\\
\delta {\Xi _{(skim)}}
\end{array} \right] = a\left[ {\begin{array}{*{20}{c}}
{{z_1}}\\
{{z_2}}
\end{array}} \right]{e^{{\lambda _{1,skim}t}}} + b\left[ {\begin{array}{*{20}{c}}
{{z_3}}\\
{{z_4}}
\end{array}} \right]{e^{{\lambda _{2,skim}t}}},\]
where $a$ and $b$ are constants (determined from the initial conditions) and $\lambda_{1,skim}$ and $\lambda_{2,skim}$ are eigenvalues of the Jacobian matrix
\begin{eqnarray}
\bf{J}_{skim} = \left( {\begin{array}{*{20}{c}}
\displaystyle{ {\left[ {\frac{{\partial U_{\perp}}}{{\partial \delta }}} \right]_{|{\delta _{s}},\Xi _{s}}}}&\displaystyle{{\left[ {\frac{{\partial U_{\perp}}}{{\partial {\Xi _p}}}} \right]_{|{\delta _{s}},\Xi _{s}}}}\\
\displaystyle{{\left[ {\frac{{\partial {\Omega}}}{{\partial \delta }}} \right]_{|{\delta _{s}},\Xi _{s}}}}&\displaystyle{{\left[ {\frac{{\partial {\Omega }}}{{\partial {\Xi _p}}}} \right]_{|{\delta _{s}},\Xi _{s}}}}
\end{array}} \right),
\label{jacobian}
\end{eqnarray}
and $\displaystyle{[z_1,~z_2]^T}$ and $\displaystyle{[z_3,~z_4]^T}$ are the eigenvectors corresponding to the two eigenvalues. The eigenvalues are given directly by: 
\begin{equation}
{\lambda _{1,skim}},{\lambda _{2,skim}} = \frac{1}{2}\left\{ {{{\left[ {\frac{{\partial {U_ \bot }}}{{\partial \delta }}} \right]}_{|{\delta _s},{\Xi _s}}} + {{\left[ {\frac{{\partial \Omega }}{{\partial {\Xi _p}}}} \right]}_{|{\delta _s},{\Xi _s}}}} \right\} \pm \frac{1}{2}{\Delta ^{1/2}},
\label{skimmingeigens}
\end{equation}
where 
\begin{equation}
\Delta  = \displaystyle{\left\{ {{{\left[ {\frac{{\partial {U_ \bot }}}{{\partial \delta }}} \right]}_{|{\delta _s},{\Xi _s}}} + {{\left[ {\frac{{\partial \Omega }}{{\partial {\Xi _p}}}} \right]}_{|{\delta _s},{\Xi _s}}}} \right\}^{2}-4\left\{ {{{\left[ {\frac{{\partial {U_ \bot }}}{{\partial \delta }}} \right]}_{|{\delta _s},{\Xi _s}}}{{\left[ {\frac{{\partial \Omega }}{{\partial {\Xi _p}}}} \right]}_{|{\delta _s},{\Xi _s}}} - {{\left[ {\frac{{\partial {U_ \bot }}}{{\partial {\Xi _p}}}} \right]}_{|{\delta _s},{\Xi _s}}}{{\left[ {\frac{{\partial \Omega }}{{\partial \delta }}} \right]}_{|{\delta _s},{\Xi _s}}}} \right\}}.
\end{equation}
Hence the stability of the skimming state will be a determined from the sign of two eigenvalues, which can both be real, or complex conjugates of each other. Once the solutions for $\delta z_{skim}$  and $\delta \Xi_{skim}$ are obtained, the solutions for $\delta x_{skim}$  and $\delta y_{skim}$ can be obtained from Eqs. \ref{xskim} and \ref{yskim}.

For the stationary state ($145^\circ < \theta_{cap} <180^\circ$), a similar analysis can be undertaken. In this case,  the inclination angle of the base state is equal to $\pi$ and the separation distance is denoted by $\delta_{f}$.  The stability is again determined by the eigenvalues of the Jacobian matrix (\ref{jacobian}), with the derivatives evaluated at $\delta_{f}$ and $\Xi_{p}=\pi$. These eigenvalues are simplified since  $\displaystyle{{{{\left[{{\partial \Omega }}/{{\partial \delta }} \right]}_{|{\delta _f},{\Xi _{s}=\pi}}}} = 0}$ and the eigenvalues are given by:
\begin{equation}
{\lambda _{1,stationary}} = {\left[ {\frac{{\partial {U_ \bot }}}{{\partial \delta }}} \right]_{|{\delta _f},{\Xi _s} = \pi }}, ~~~~~{\lambda _{2,stationary}} = {\left[ {\frac{{\partial \Omega }}{{\partial {\Xi _p}}}} \right]_{|{\delta _f},{\Xi _s} = \pi }},
\label{stationaryeigens}
\end{equation}
Therefore, in the stationary state the two eigenvalues are real. 

For both states determining the stability is computationally straightforward. For a given active area $\theta_{cap}$, for the skimming state, the separation distance and inclination angle are computed, and for the stationary state, just the separation distance is obtained. The separation distance  and inclination angle is then changed over small values from the base state, and the velocity in the $z$ direction and the rotation are computed to obtain the derivatives of the Jacobian matrix.  Once these derivatives are obtained, the eigenvalues are computed straightforwardly.  To undertake this calculation and evaluate partial derivatives a fourth order central finite difference scheme is used with interval width or grid spacing $\Delta$ given by $\Delta\delta =0.01$ and $\Delta \Xi =1.0^{\circ}$. Changing this grid spacing to half did not change the eigenvalues reported in tables \ref{sk1}, \ref{si2} to within a few percent.

\subsection{Brownian Dynamics Simulation and Algorithm}
\label{brownianformulation}
At the continuum level,  the colloidal locomotor moving in the continuous liquid phase  is subject to the deterministic diffusiophoretic propulsive forces and torques $\displaystyle{({\bf{\tilde F}}_P,~{\bf{\tilde T}}_P)^T}$
(Eq. \ref{propulsiveforcetorque1} and \ref{propulsiveforcetorque2}), 
hydrodynamic drag forces and resistive torques  $\displaystyle{({\bf{\tilde F}}_H,~{\bf{\tilde T}}_H) ^T(= - \bm{\mathcal{\tilde R}} ({{\bf{\tilde U}}},~{{\bf{\tilde \Omega}}})^T }$,
Eq. \ref{hydrodynamicresistance}), 
and fluctuating Brownian forces and torques $\displaystyle{\bm{\mathcal{\tilde{B}}} = ({\bf{\tilde F}}_B,~{\bf{\tilde T}}_B)^T}$, 
 which represent the effects of (unresolved) sub-micron scale atomic interactions with fluid molecules. The Brownian force is characterized by its moments,
\begin{gather}
\left\langle \bm{\mathcal{\tilde{B}}}(\tilde t)\right\rangle=0,\\
\left\langle \bm{\mathcal{\tilde{B}}}(\tilde t) ~\bm{\mathcal{\tilde{B}}}(\tilde{t'}) \right\rangle=2k_BT~ \bm{\mathcal{\tilde R}} ~\delta(\tilde t - \tilde t' ), 
\label{brfo}
\end{gather}
where $\langle\cdot\rangle$ denotes an ensemble average, $k_B$ is Boltzmann's constant and $T$ is the fluid temperature. Eq.~\ref{brfo}, which 
relates the strength of Brownian and hydrodynamic forces, is a consequence of the fluctuation-dissipation theorem and ensures that the system is in
local thermodynamic equilibrium. 

The equation of motion is the Langevin stochastic differential equation \cite{brady1988}, which we simplify to the overdamped limit because of the assumption of low Reynolds number flow:
\begin{equation}
\left( {\begin{array}{*{20}{c}}
{\bf{\tilde F}}_P\\
{\bf{\tilde T}}_P
\end{array}} \right) 
+\left( {\begin{array}{*{20}{c}}
{\bf{\tilde F}}_H\\
{\bf{\tilde T}}_H
\end{array}} \right) 
+\left( {\begin{array}{*{20}{c}}
{\bf{\tilde F}}_B\\
{\bf{\tilde T}}_B
\end{array}} \right)
=0,
\label{brfo1}
\end{equation}
Following standard methods \cite{ermak1978}, the Langevin equation can be integrated in time to yield an equation for the
particle displacements in the overdamped limit:
\begin{equation}
\Delta {\bf{\tilde x}} = {\bf{\tilde M}}\cdot{\bm{\mathcal{\tilde F}}}_P\, \Delta \tilde t + k_BT\,\bm{\tilde \nabla}\cdot{\bf{\tilde M}}\,\Delta \tilde t + \sqrt{2k_BT}\,\bf{\tilde B}\cdot\Delta {\bf{\tilde w}}, 
\label{ito}
\end{equation}
where ${\bf{\tilde x}}$ is a six-dimensional vector of particle translational and angular displacements, $\bf{\tilde M}=\tilde {\bm{\mathcal{R}}}^{-1}$ is the mobility matrix, $\displaystyle{\bm{\mathcal{\tilde F}}_{p} = ({\bf{\tilde F}}_P,~{\bf{\tilde T}}_P) ^T}$ and 
${\bf{\tilde B}}$ is a lower triangle matrix satisfying $\bf{\tilde B}\cdot\bf{\tilde B}^T = {\bf{\tilde M}}$ which can be obtained form a Cholesky decomposition of $\bf{\tilde M}$ (note $\bf{\tilde M}$ is a symmetric positive-definite matrix).  The last term involves $\Delta {\bf{\tilde w}}$, an Wiener process whose first and second moments are $\langle\Delta\bf{\tilde w}\rangle=0$ and 
$\langle\Delta \tilde w_i\;\Delta \tilde w_j\rangle=\delta_{ij}\,\Delta \tilde t$, in practice six independent Gaussian random numbers.
The equation is written in the It\^{o} convention where the mobility matrix elements and forces/torques are computed at the beginning of each time step. The second term on the right hand side of Eq.~\ref{ito} is the thermal drift, which arrises from a simple forward Euler integration scheme of the Langevin equation for the case in which the resistance tensor varies spatially. Alternatively, higher order integration schemes (e.g. the midpoint algorithm \cite{fixman1978}) can be used which result in equations which do not  require computation of the thermal drift term. 
\section{Results}
\subsection{Linear Hydrodynamic Stability}
We begin with the stability of the skimming state (Fig. \ref{schematic2}a), which is realized for active areas for which $115^{\circ}<\theta_{cap}<145^{\circ}$. The eigenvalues are given by Eqn. \ref{skimmingeigens}. We choose six active cap angles (see Table \ref{sk1}) which correspond to different steady separation distances and inclination angles ($\delta_{s}, \Xi_{s}$), and compute the eigenvalues as given in Table \ref{sk1}. We note that for all the states, the real part of the eigenvalues are negative, indicating that all these stationary states are \textit{stable}. In addition we note that oscillatory behavior is obtained for the higher range of the active cap areas.
\begin{table}[ht] 
\centering 
\begin{tabular}{c c c c c }
 \hline\hline $\theta_{cap}$ & $\delta_{s}$ & $\Xi_{s}$ & $\lambda_{1,skim}$& ~~~$\lambda_{2,skim}$\\ [0.5ex]	
  \hline 
$115^{\circ}~$&~$0.359$~~~&~$123.4^{\circ}$~~~&~$-0.161916$~~~&~$-0.029100$\\
$120^{\circ}~$&~$0.243$~~~&~$132.8^{\circ}$~~~&~$-0.155265$~~~&~$-0.100353$\\
$125^{\circ}~$&~$0.177$~~~&~$141.0^{\circ}$~~~&~$-0.124352 - 0.118331~i$~~~&~$-0.124352 + 0.118331~i$\\
$130^{\circ}~$&~$0.133$~~~&~$148.7^{\circ}$~~~&~$-0.142354 - 0.163835~i$~~~&~$-0.142354 + 0.163835~i$\\
$135^{\circ}~$&~$0.102$~~~&~$156.3^{\circ}$~~~&~$-0.128264 - 0.197313~i$~~~&~$-0.128264 + 0.197313~i$\\
$142^{\circ}~$&~$0.071$~~~&~$168.5^{\circ}$~~~&~$-0.073978 - 0.167182~i$~~~&~$-0.073978 + 0.167182~i$\\
  \end{tabular} 
  \caption{Eigenvalues for the stability of the  skimming state for various active area coverages, $i=\sqrt{-1}$.} 
  \label{sk1} 
  \end{table} 

 We repeat the computation for the stationary state (Fig. \ref{schematic2}a) which occurs for larger active areas $145^{\circ}<\theta_{cap}<180^{\circ}$. In this case, as noted in the derivation, the eigenvalue are all real (Eqn. \ref{stationaryeigens}) and we obtain (see Table \ref{si2}) the eigenvalues for six different cap sizes and find that the stationary state is also stable.
\begin{table}[ht] 
\centering 
\begin{tabular}{c c c c c }
 \hline\hline $\theta_{cap}$ & $\delta_{f}$ & $\Xi_{f}$ & $\lambda_{1,stationary}$& ~~~$\lambda_{2,stationary}$\\ [0.5ex]	
  \hline 
$145^{\circ}~$&~$0.065$~~~&~$180^{\circ}$~~~&~$-0.093001$~~~&~$-0.008476$\\
$147^{\circ}~$&~$0.22$~~~&~$180^{\circ}$~~~&~$-0.098059$~~~&~$-0.058091$\\
$150^{\circ}~$&~$0.38$~~~&~$180^{\circ}$~~~&~$-0.070108$~~~&~$-0.043372$\\
$152^{\circ}~$&~$0.55$~~~&~$180^{\circ}$~~~&~$-0.029184$~~~&~$-0.024499$\\
$155^{\circ}~$&~$0.83$~~~&~$180^{\circ}$~~~&~$-0.054003$~~~&~$-0.015050$\\
  \end{tabular} 
  \caption{Eigenvalues for the stability of the stationary state for various active area coverages.} 
  \label{si2} 
  \end{table} 

\subsection{Brownian Dynamic Study of the Skimming and Stationary States}
As reviewed earlier, in the deterministic limit in which Brownian forces are absent, Mozaffari \emph{et al.} \cite{mozaffari2016} formulated the solute conservation (mass transfer) equation and the continuity and Stokes equations for the diffusiophoretic slip velocity in bispherical coordinates, and by analytically solving these equations obtained the active (diffusiophoretic) propulsive forces and torques on the locomotor, 
$\displaystyle{({\bf{\tilde F}}_P,~{\bf{\tilde T}}_P)^T}$
(see also Sec. \ref{propulsiveformulation}). From the requirement that the locomotor is force and torque free, the deterministic velocity $\bf{{\tilde U}}$ and rotation $\bm{\tilde \Omega}$ vectors of the locomotion are computed  by summing these active forces and torques with the hydrodynamic resistive viscous forces and torques ,
$\displaystyle{({\bf{\tilde F}}_H,~{\bf{\tilde T}}_H)^T}$
to zero, where the viscous resistance is given by $ - \bm{\mathcal{\tilde R}}~ ({{\bf{\tilde U}}},~{{\bf{\tilde \Omega }}})^T$
(see also Sec. \ref{resistiveformulation}) and Eq.~\ref{Rmatrix} for $\bm{\mathcal{\tilde R}}$. The particle translates in the plane formed by the orientation axis $\bm{n}_{p}$ and the wall normal vector $\bm{\hat e}_z$ and it simultaneously rotates about an axis parallel to the cross product of the latter two vector. Mozaffari \emph{et al.} \cite{mozaffari2016} obtained, for a range of active areas ($\theta_{cap}$), the translational velocities perpendicular and along the surface  ($U_{\perp}(\delta, \Xi;\theta_{cap})$ and $U_{\parallel}(\delta, \Xi;\theta_{cap})$, both scaled  by $\tilde v_{c}$)  and the   rotation rate  ($\Omega(\delta, \Xi;\theta_{cap})$, scaled by $\tilde v_{c}/a$). From the translational and rotational velocities, locomotor  trajectories were computed. It was shown that the general nature of the trajectory of a particle near the wall is related to its initial orientation and the angular coverage of the active cap.  If the initial orientation of the active area faces the wall,  the accumulation of solute in the gap between the wall and the locomotor  drives the locomotor  away from it. On the other hand, when the cap faces opposite to the wall, the particle moves towards the wall (Fig. \ref{schematic2}).  Regardless of the coverage size, when initially  $\Xi<90^\circ$ the particle is repelled from the wall. However for larger initial $\Xi$ (initial angle of incidence), the solute build up on the outward side initially drives the particle towards the wall.  If the coverage is not too large ($\theta_{cap}<115^\circ$), the particle scatters from the wall while for a range of higher coverage $115^\circ < \theta_{cap} < 145^\circ$, the particle skims along the wall at a constant separation $\delta_{s}$ with constant  orientation $\Xi_{s}$ (Fig. \ref{schematic2}a). For very high coverage $\theta_{cap} > 145^\circ$ the particle moves to a stationary state very close to the wall with zero translational and angular velocity  while holding a fixed orientation $\Xi_{eq}=180^\circ$ (Fig. \ref{schematic2}b).  

In this section we describe the influence of Brownian forces on the skimming and stationary regimes of boundary guidance by adding the Brownian fluctuation force to the force balance (see Eq.~\ref{brfo1}). For the numerical computations we write the evolution equation Eq.~\ref{ito} in dimensionless form as 
\begin{equation}
\Delta {\bf{x}} = {\rm P}e\,(\bm{\mathcal{R}} ^{ - 1} \cdot {\bm{\mathcal{F}}}_P \Delta t) + \bm{\nabla} \cdot \bm{\mathcal{R}} ^{ - 1} \,\Delta t + 
\sqrt 2 \,{\bf{B}}\cdot\,\Delta {\bf{w}}, \label{ito2}
\end{equation}
where the colloid P\'eclet number is defined as ${\rm P}e = \pi \mu a^2 \tilde v_c/k_B T$. In this equation, ${\bf{x}} = {\bf{\tilde x}}/a$, $\Delta t = \Delta \tilde t/t^*$, and
$\Delta {\bf{w}} = \Delta {\bf{\tilde w}}/\sqrt{t^*}$ where $t^*=\pi\mu a^3 /k_B T$ is the rotational diffusion time. Notice for $\bm{\mathcal{F}}_P$,  the force and torque entries are 
non-dimensionalized with $\pi\mu a\tilde v_c$ and $\pi\mu a^2\tilde v_c$ respectively, and ${\bf{B}}\cdot{\bf{B}}^T  = \bm{\mathcal{R}}^{-1}$. At each time step,  care is needed when the orientation of the particle is updated since rotations about different axis do not commute. Here we use  an unbiased rotational moves algorithm \cite{beard2003}. (This reference contains a typographical error --  the (2,2) element of the transformation matrix Eq.~5 should read $[(\Omega_a^2+\Omega_c^2)\,\cos\Omega +\Omega_b^2]/\Omega ^2$.)

\begin{figure} [!htb]
\centering
\includegraphics[width=0.50 \textwidth]{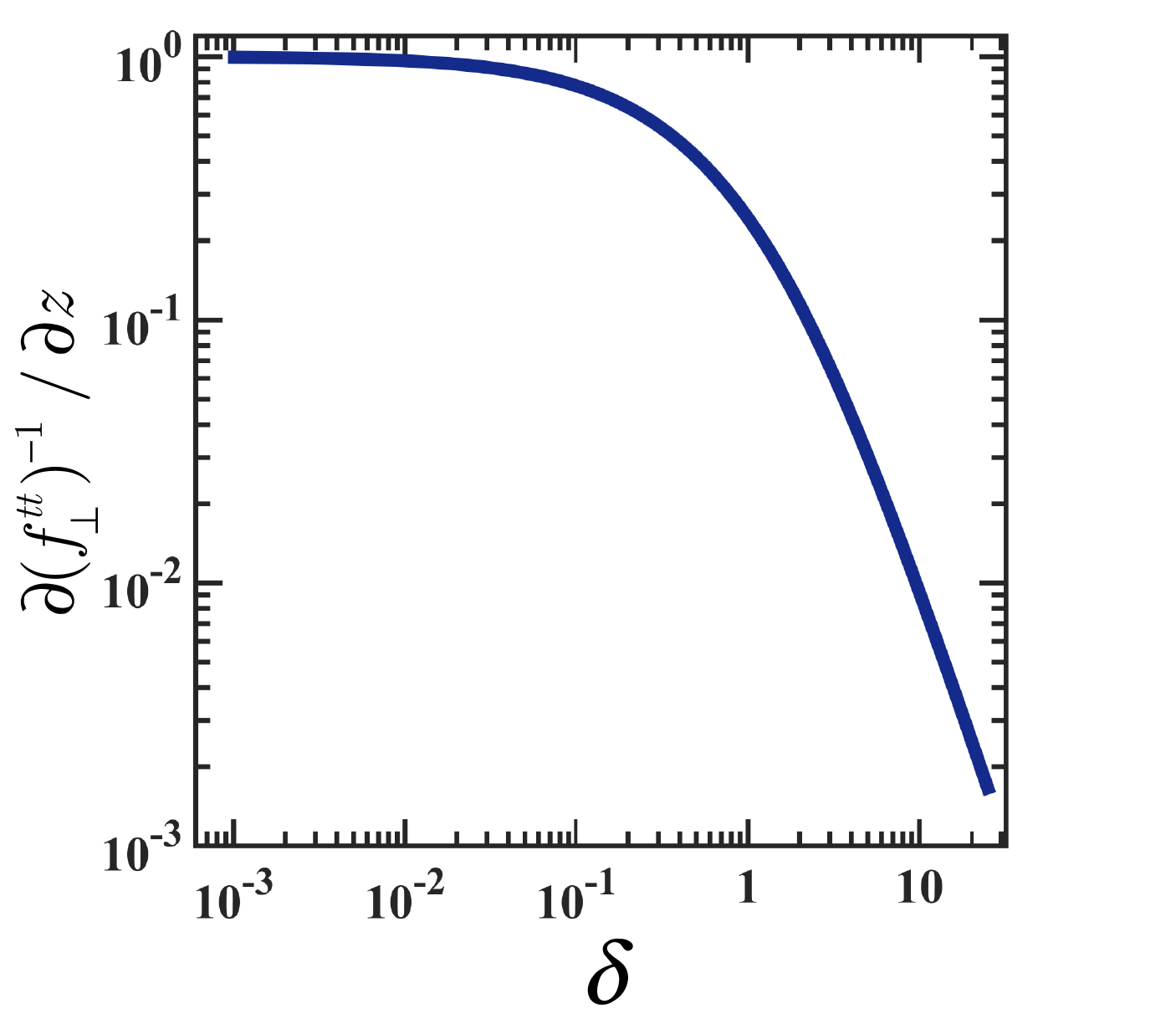}     	    
\caption{\footnotesize{Thermal drift term}} 
\label{deldm}
\end{figure}

The mobility is only a function of the distance to the wall and becomes constant at large separation, so the second ``drift'' term on the right hand side of 
Eq.~\ref{ito2} (which arises in the passage to the overdamped limit) has only a $z$-component and acts to push particles away from the wall \cite{lau2007}.  This term appears from
the $f^{tt}_\bot$ entry of the resistance matrix, and its contribution to the equation of motion is illustrated in Fig.~\ref{deldm}.

In the following, to obtain  smooth and yet statistically significant results, all statistical outputs including probability distribution, mean square displacement and average displacement along or out of the initial skimming direction were found by utilizing approximately $25,000$ various realizations.
\subsubsection{Skimming Swimmers}
\begin{figure} [!htp]
\centering
\includegraphics[width=0.75 \textwidth]{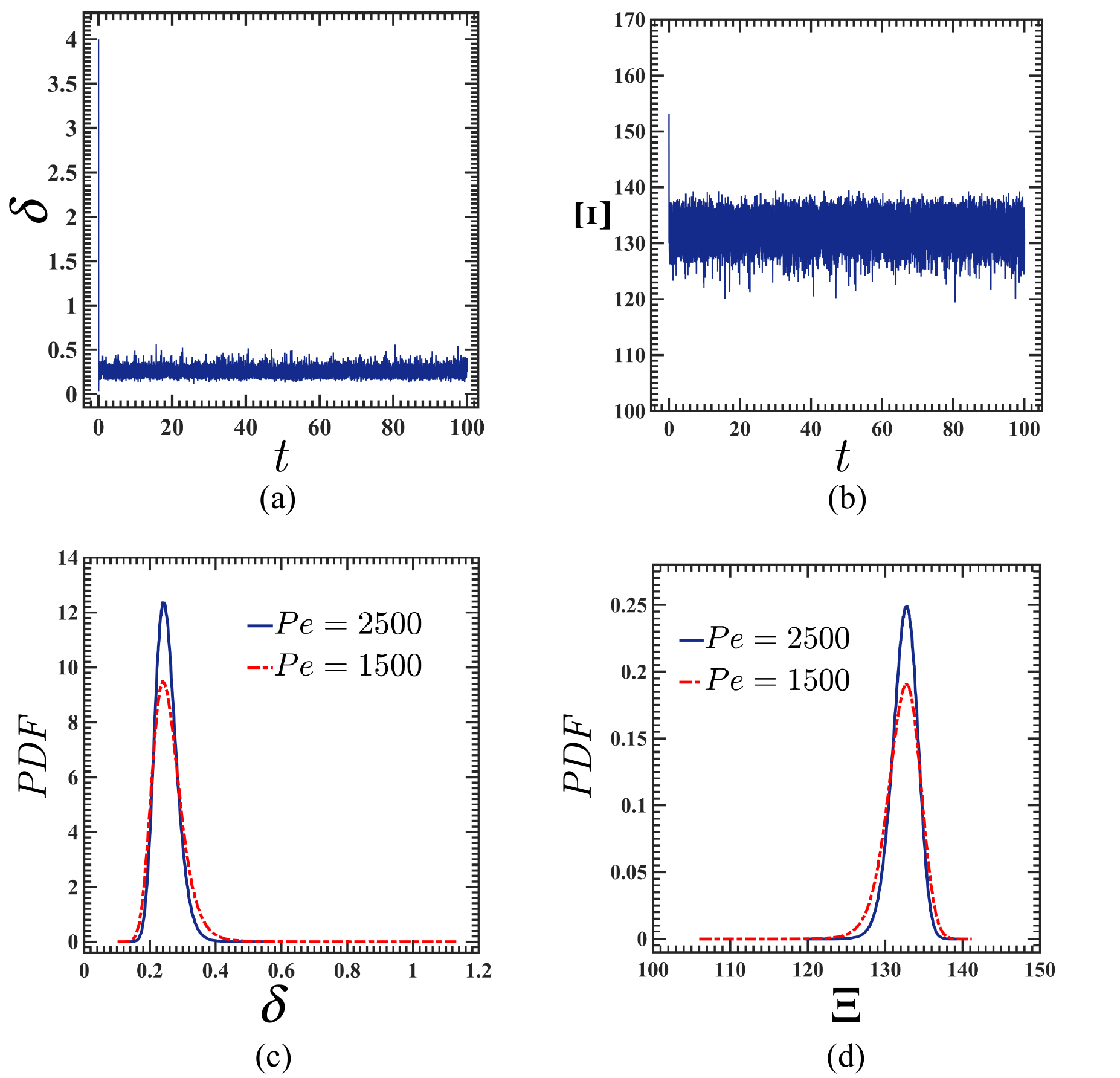}     	    
\caption{\footnotesize{Swimmer with $\theta _{cap}=120^\circ$ and initial configuration $\delta=4.0$ and $\Xi=150^\circ$ at ${\rm P}e=1500$. (a) Variation of gap with time, (b) variation of tilt angle with time (single realization), (c) Probability distribution of gaps for ${\rm P}e=1500$ and 2500, (d) Probability distributions of tilt angle (obtained from a very long time simulation).}} 
\label{skim1}
\end{figure}

\begin{figure} [!htb]
\centering
\includegraphics[width=0.80 \textwidth]{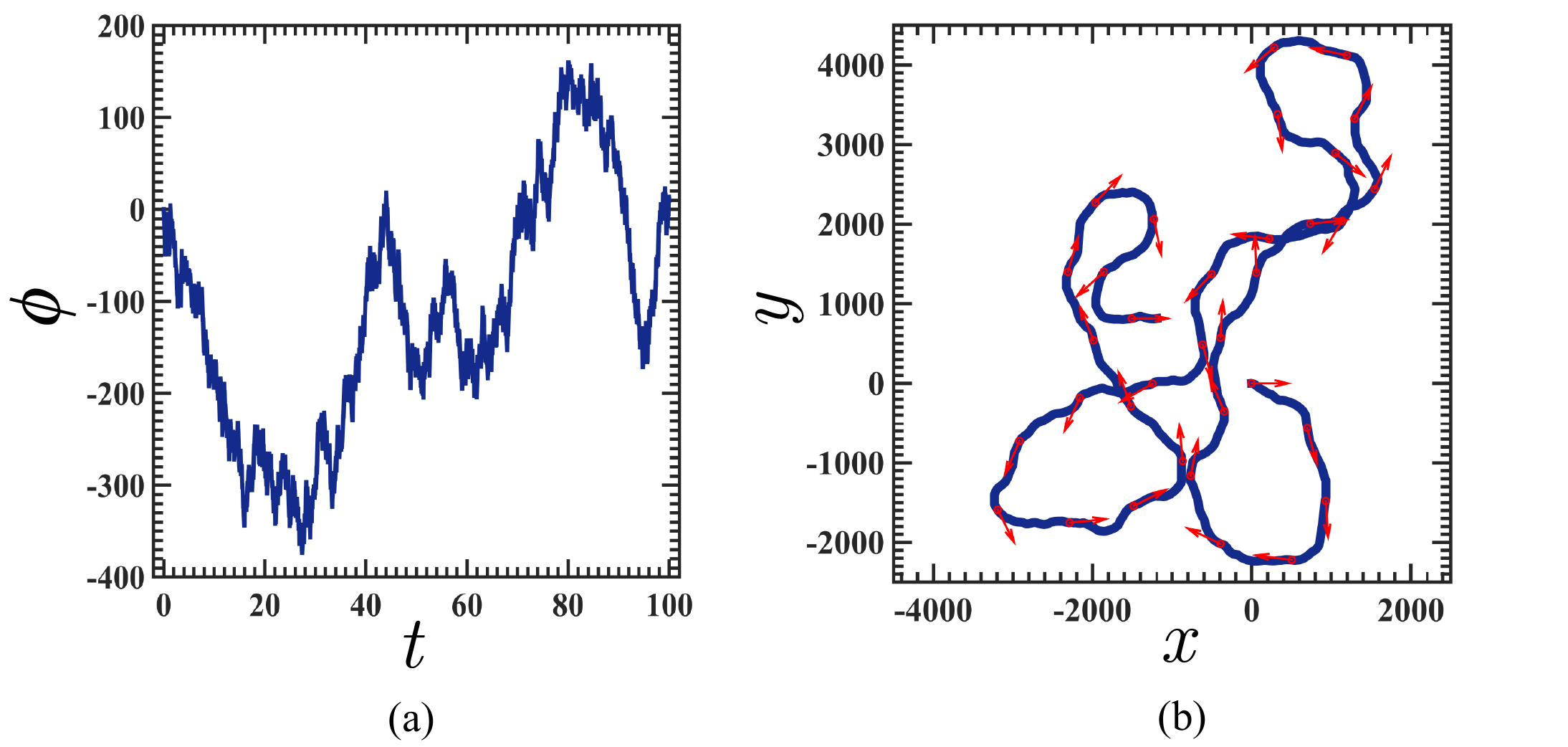}     	    
\caption{\footnotesize{Same swimmer as in Fig.~\ref{skim1}. (a) Variation of the in-plane orientation angle $\phi$ with time, (b) Trajectory in the $x$-$y$ plane. The red arrows represent the projection of orientation vector of particle on the $x$-$y$ plane.}} 
\label{skim2}
\end{figure}

Normally at high P\'eclet number, e.g. ${\rm P}e > 1000$, one expects small fluctuations about a deterministic trajectory but here small changes in orientation can have drastic effects
on the motion. As shown in Fig.~\ref{skim1}, the gap and the tilt angle $\Xi$ are quite stable in this limit, with probability distributions that peak sharply about the deterministic 
values $\delta_{s}=0.22$ and $\Xi_{s}=133^\circ$ while broadening as ${\rm P}e$ decreases. The fact that $\Xi$ is stable is a necessary requirement for skimming at nearly-constant 
$\delta$.  However, there are also fluctuations in $\phi$, the angle the projection of the orientation vector $\bm{n}_{p}$ onto the wall makes with the $x$-axis. In Fig. \ref{skim2}a we plot $\phi$ as a function of time $t$, and note that the projection of the orientation vector onto the $x-y$ plane rotates by $2\pi$ over 20 time units, and then reverses its rotation returning to its original projection direction after another approximately 20 time units. Hence the locomotor is constantly rotating around the $z-axis$ as it maintains a tilt of $\Xi_{s}$ off of the axis. This rotation about the $z$ axis is also evident in Fig. \ref{skim2}b which shows the direction of the projection of the orientation vector in the $x-y$ plane as the locomotor moves along this plane while maintaining a mean position $\delta_{s}$ above the plane.  The rotation about the $z$ axis is the result  of the Brownian  torque about this axis and results in a persistent random walk in the $x$-$y$ plane.
\begin{figure} [!htb]
\centering
\includegraphics[width=0.80 \textwidth]{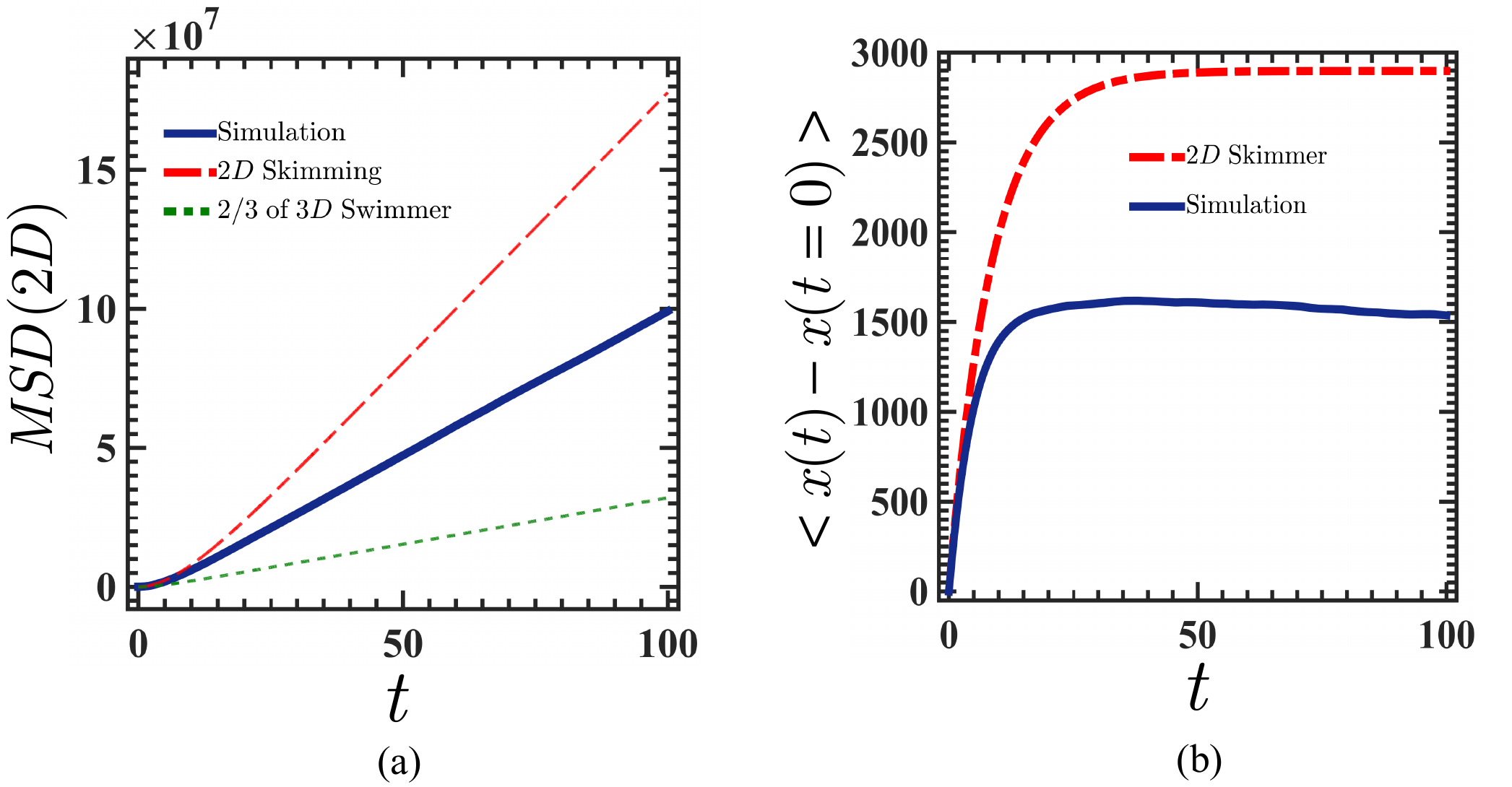}     	    
\caption{\footnotesize{Moments of the particle displacement at ${\rm P}e=1500$. (a) Mean square displacement for the simulated swimmer (blue), 2d skimmer (red) and a swimmer in an infinite medium (green).  (b) Average $x$-position of the simulated swimmer and the 2d skimmer.}} 
\label{skimmsd}
\end{figure}
\begin{figure} [!htb]
\centering
\includegraphics[width=0.70\textwidth]{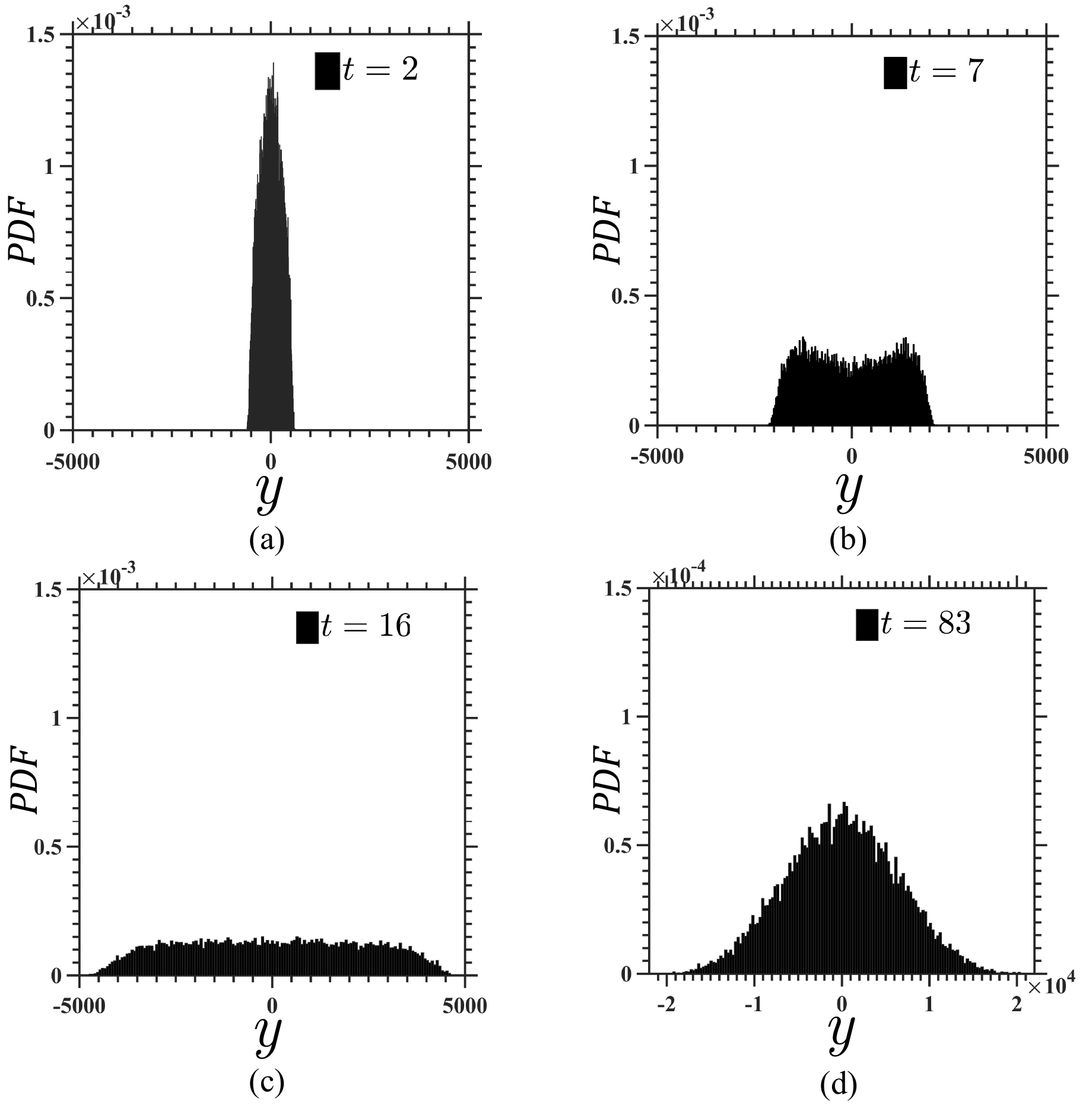}     	    
\caption{\footnotesize{Probability distributions of the $y$-position of a simulated swimmer at different times for ${\rm P}e=1500$.}} 
\label{pdfy}
\end{figure}

We wish to further analyze the in-plane random motion of the skimmers, in the limit of high ${\rm P}e=1500$ where skimming is persistent.  We first compare the mean-square displacement
(MSD) of a such skimmer against two different cases: $(a)$ a hypothetical skimmer moving purely in two dimensions at the deterministic equilibrium values of gap and tilt ($\delta_{s}$, $\Xi_{s}$) but with 
rotational thermal noise acting on the particle only in the $z$-direction and $(b)$ 2/3 of the MSD of an unconfined three-dimensional swimmer. We see in Fig.~\ref{skimmsd}a that
in all cases there is a transition from ballistic to diffusive motion at times distinctly longer that the rotational diffusion time, and that in the latter regime the three 
situations have distinctly different diffusivities (slopes). The difference between the purely two and three dimensional swimmers is that in 2d the orientation vector moves on 
a unit circle and decays exponentially with time as $\langle\hat u(t)\cdot\hat u(0)\rangle = \exp(-D_Rt)$,  slower than a 3d swimmer where the orientation vector is diffuses 
on the unit sphere with correlation function decaying as $\langle\hat u(t)\cdot\hat u(0)\rangle = \exp(-2D_Rt)$ (see Appendix~\ref{append2}). The orientation of the 2d skimmer remains correlated for a longer 
time so uni-directional motion is more persistent. The simulated swimmer is in between since its orientation vector is not completely free to diffuse on the unit sphere but has 
more freedom in comparison with the 2d skimmer. A similar trend is apparent in Fig.~\ref{skimmsd}b which compares the average displacement along the initial skimming direction 
$\langle x(t)-x(0)\rangle$ for purely 2d skimmers and the simulated swimmer.  In both cases, rotational diffusion eventually randomizes the direction of motion, but in the former 
case the particle moves further on average since its orientation is more persistent. A similar comparison can be made for the average $y$-displacements but more interestingly,
the probability distributions for simulated swimmers, Fig.~\ref{pdfy}, shows non-Gaussian behavior initially, followed by a transition to a Gaussian shape at long times in the 
diffusive regime.
 \begin{figure} [!htb]
\centering
\includegraphics[width=0.95 \textwidth]{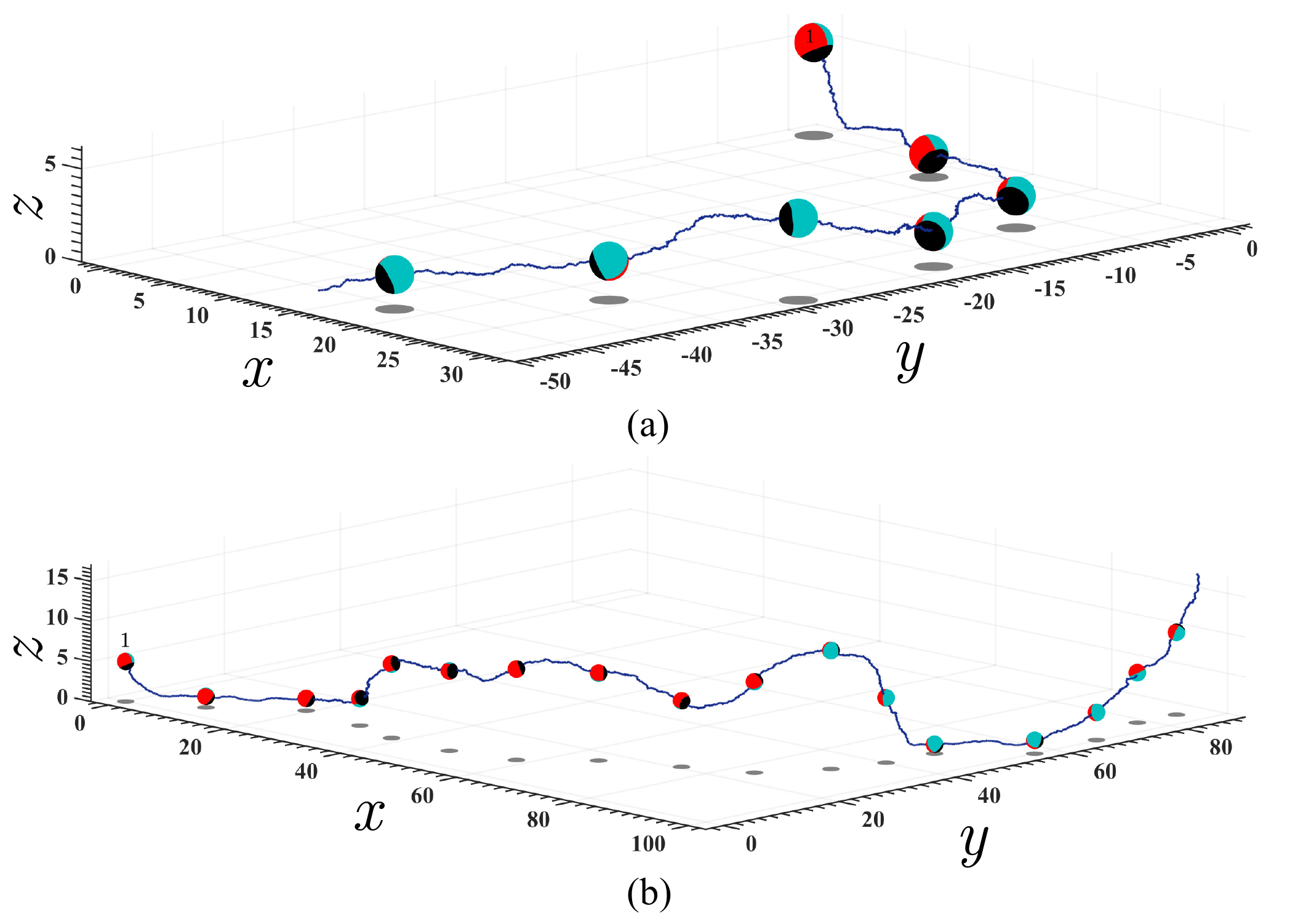}     	    
\caption{\footnotesize{Two sample trajectories for swimmer with $\theta _{cap}  = 120^ \circ$ and an initial configuration $\delta=4.0$, $\Xi=150^ \circ$ for ${\rm P}e=50$. The black color shows the passive part. The active side represented by red and light blue colors to better show the rotation.}} 
\label{skim50}
\end{figure}

When the P\'eclet number decreases further the fluctuations in tilt and gap become more pronounced, and only transient skimming states are observed once ${\rm P}e<200$. Two typical 
trajectories are given in Fig.~\ref{skim50}, the second illustrating the possibility of a particle skimming, leaving the wall region, and returning to skim in a different direction.   A useful characterization of the skimming states is the persistence time, the interval over which skimming persists, as a function of P\'eclet number. The average time is not
the most suitable measure since the calculations have a time limit and in some cases skimming does not terminate, so instead we give the probability distributions of skimming 
\begin{figure} [!htb]
\centering
\includegraphics[width=0.75 \textwidth]{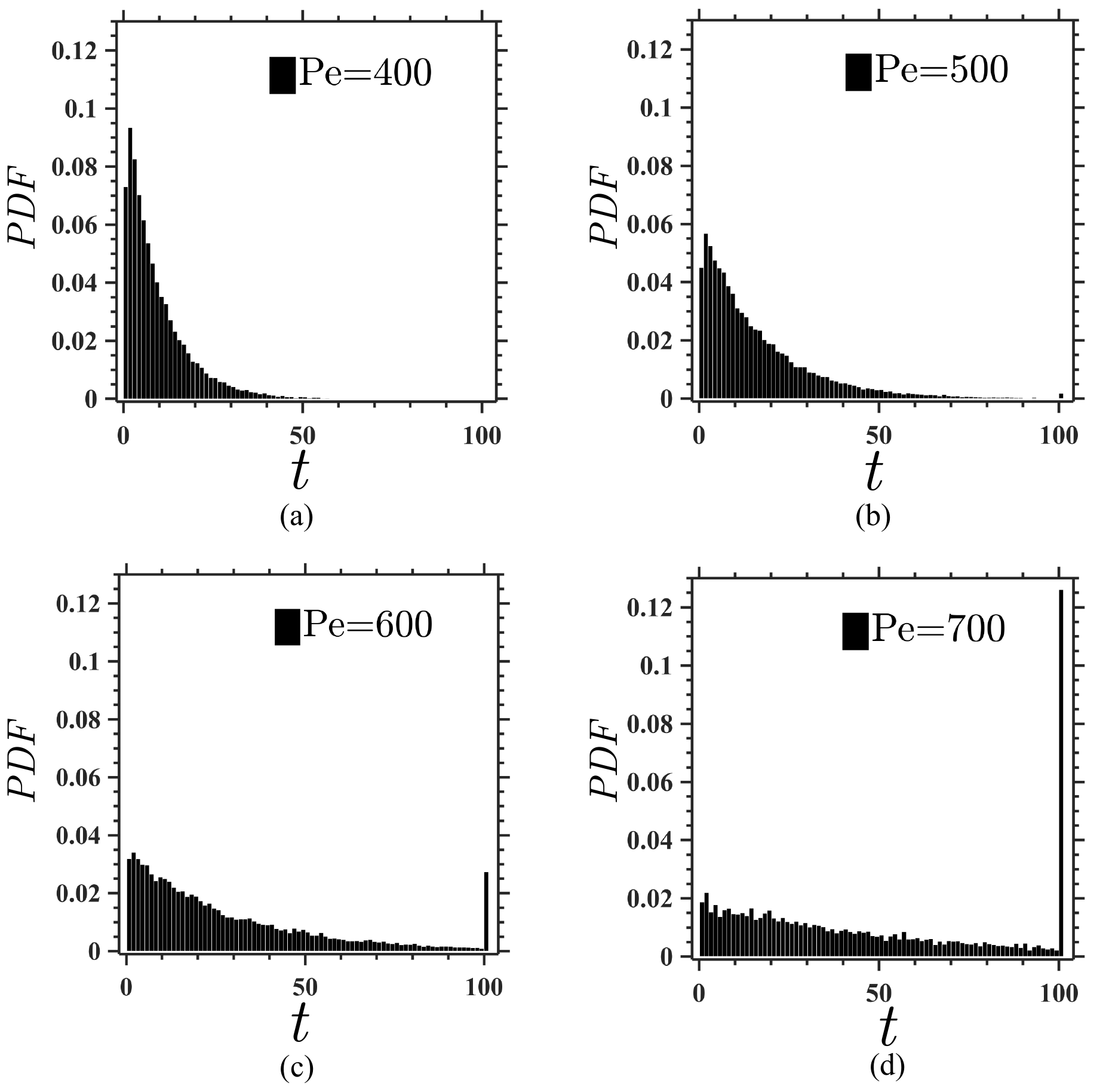}     	    
\caption{\footnotesize{Probability distribution of skimming time for four different values of the P\'eclet number.}} 
\label{skimhist}
\end{figure}
time for different (moderate) values of ${\rm P}e$ in Fig.~\ref{skimhist}.  More precisely, we define the skimming time as the first passage interval between a particle reaching 
$\delta_{eq}$ and $\delta_{eq}+1$. We have also used other criteria such as the time at which the tilt angle becomes less than $90^\circ$ but the results have the same trends and
general behavior. We have seen in the figure that as the P\'eclet number increases the skimming time distribution shifts toward longer values, and by ${\rm P}e=700$ most of the particles 
do not leave the plane for the duration of the simulation. 
\subsubsection{Stationary swimmers}
\begin{figure} [!htb]
\centering
\includegraphics[width=0.95\textwidth]{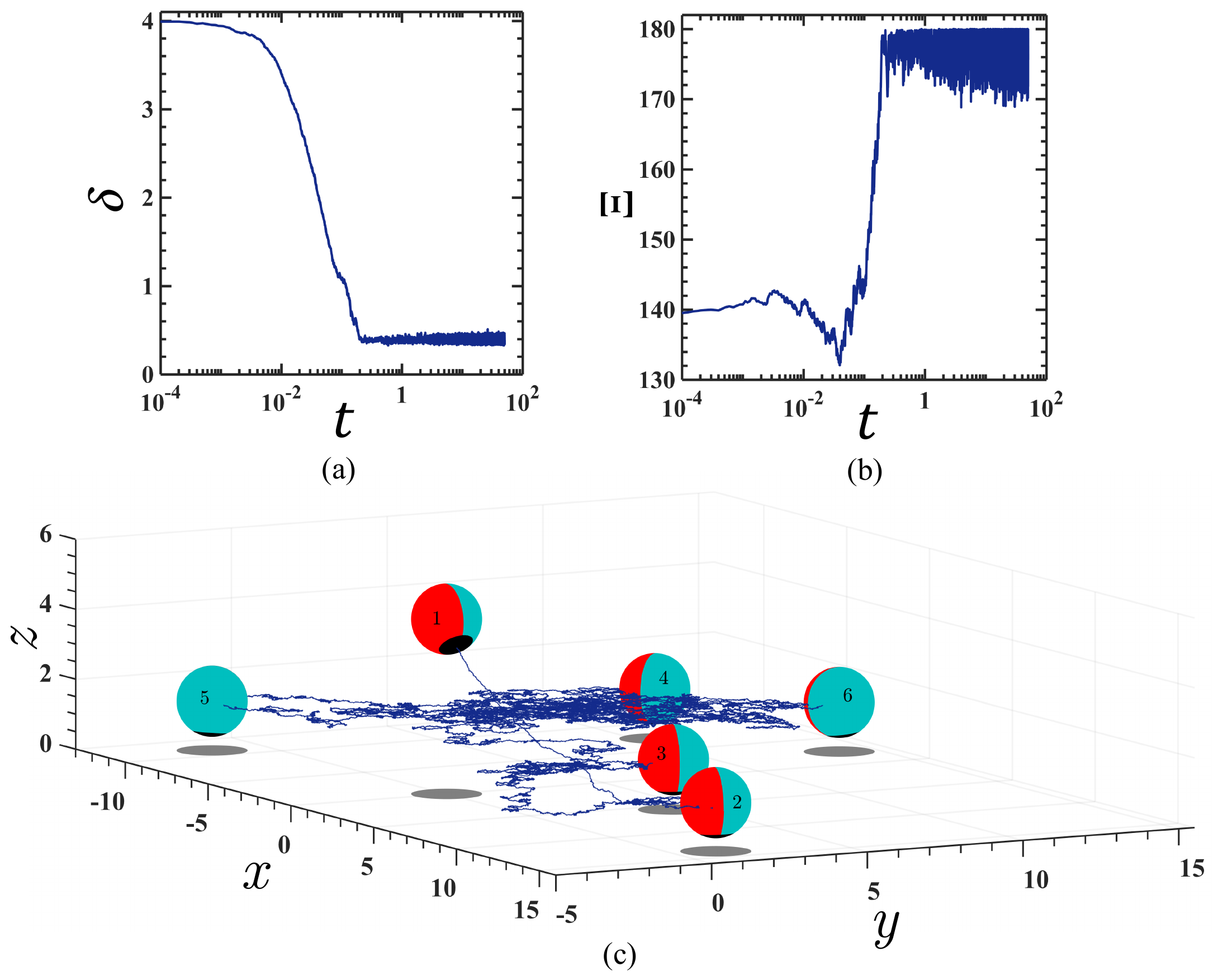}     	    
\caption{\footnotesize{Example of a swimmer with $\theta _{cap}=150^\circ$ and initial configuration $\delta=4.0$ and $\Xi=140^\circ$ for ${\rm P}e=1500$. (a) Variation of separation distance with time, (b) Variation of tilt angle with dimensionless time and (c) A sample trajectory, indicating the orientation.}} 
\label{stattr}
\end{figure}
\begin{figure} [!htb]
\centering
\includegraphics[width=0.90\textwidth]{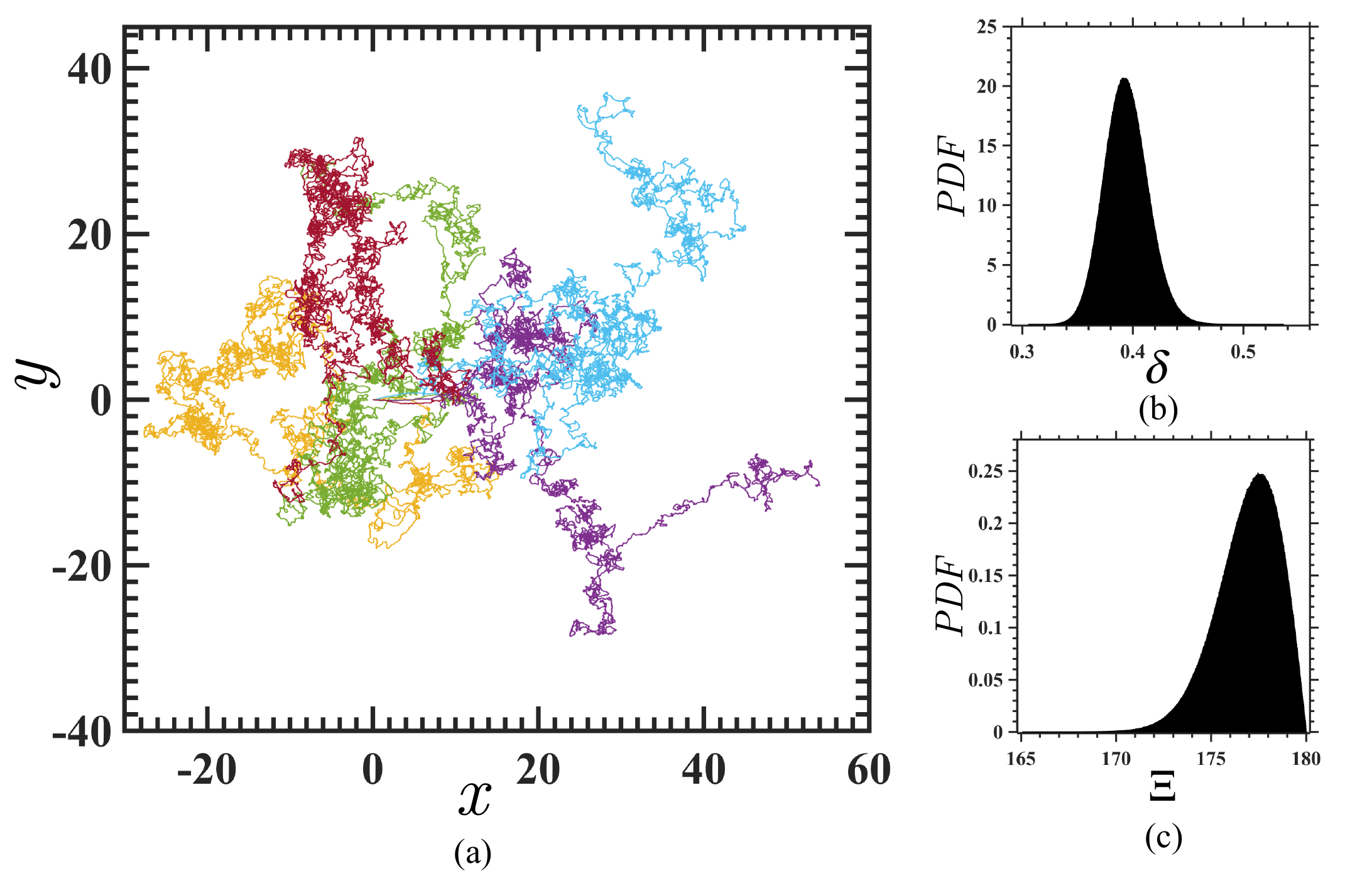}     	    
\caption{\footnotesize{Swimmers at ${\rm P}e=1500$ with initial condition $\delta=4.0$ and $\Xi=140^\circ$ (a) Projection of trajectories in the $x$-$y$ plane for five realization, (b) PDF of gap distance and (c) PDF of inclination angle.}} 
\label{sch}
\end{figure}

In the absence of thermal fluctuations, a particle with very high coverage $\theta_{cap}>145^\circ$ and an initial orientation where the particle cap is opposite to the wall comes to a stop at the equilibrium
gap distance $\delta_{f}$ and a tilt angle $\Xi_{f}=180^\circ$, Fig. \ref{schematic2}b. When the P\'eclet number is large, ${\rm P}e > 1000$, this remains a stable state: the gap and 
orientation fluctuate about the deterministic values, as shown in Fig.~\ref{stattr} for a typical example. Fluctuations in $\delta$ are weak because the stationary position is an 
attractor due to the opposing propulsion and lubrication forces there. Fluctuations in $\Xi$ produce a force parallel to the wall which is resisted by lateral hydrodynamic drag, 
resulting in somewhat larger motions parallel to the wall. The in-plane rotation is in random directions due to fluctuations in $\phi$ which are unsurpressed because the deterministic
forces does not give torque about the $z$-axis.  Going beyond this single example, Fig.~\ref{sch} shows the lateral motions for five sample realizations at ${\rm P}e=1500$, along with the 
probability distribution functions of gap and inclination angle.  

\begin{figure} [!htb]
\centering
\includegraphics[width=0.75\textwidth]{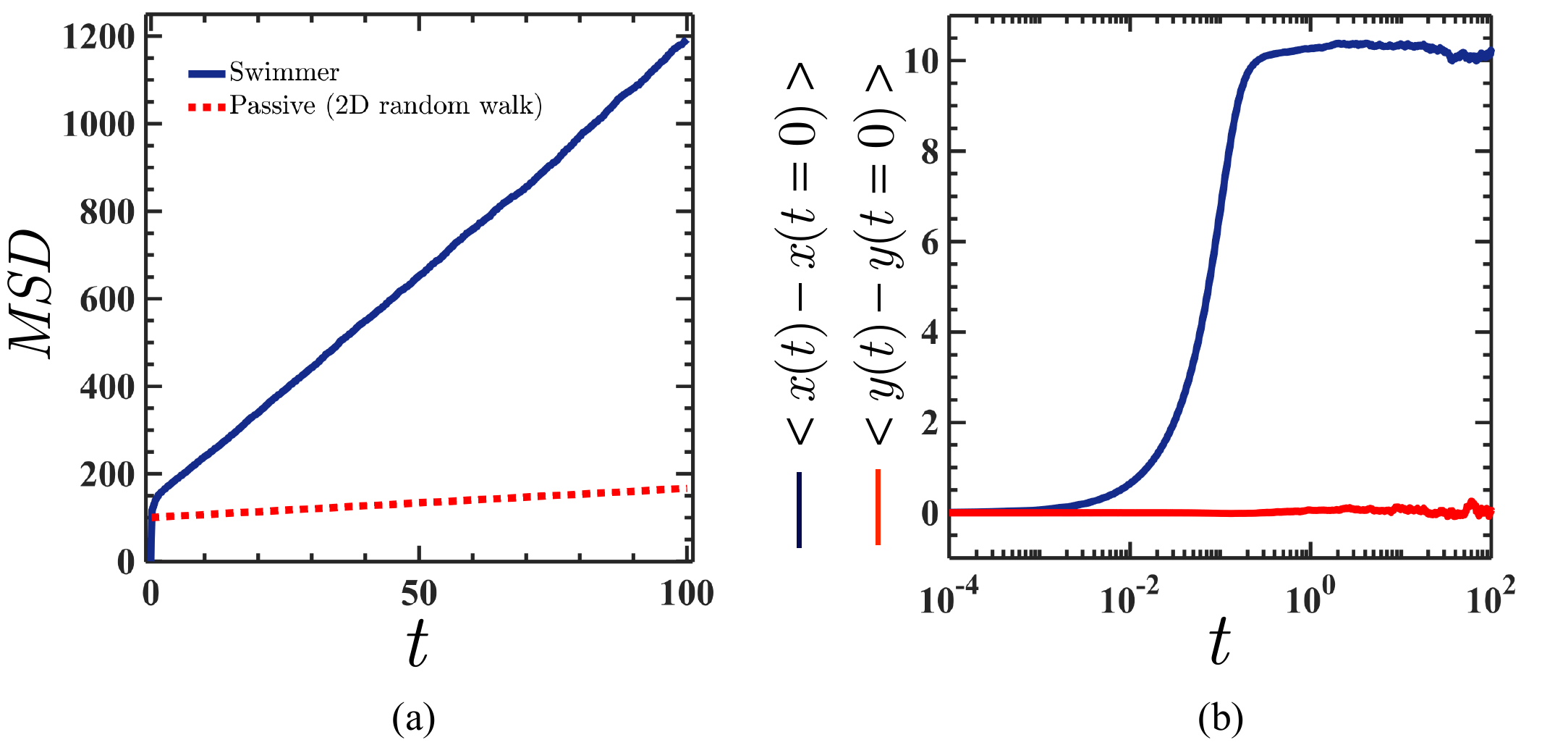}     	    
\caption{\footnotesize{Swimmer at ${\rm P}e=1500$ approaching the stationary configuration. (a) In-plane mean square displacement {\sl vs}. time for the swimmer (solid line) compared to a passive 2d random walker (dashed line),  (b) average displacement along $x$ and $y$ axis as a function of time.}} 
\label{msdstat}
\end{figure}

The diffusive motion can be characterized by the mean square displacement, shown in Fig.~\ref{msdstat}a, which indicates an effective diffusivity much greater than that of an ordinary 
two dimensional random walker with the same molecular diffusion coefficient.  The reason for the enhancement is that fluctuations in the tilt angle away from $180^\circ$ lead to
a non-zero propulsion force which has a transient persistence in direction before decorrelating.  The rapid increase at early times appears in the figure because $t=0$ 
corresponds to the initial state of the simulations, when the particle is placed away from the surface and translates before settling down to its stationary configuration.  This
behavior is more evident in the plot of the displacement {\sl vs}. time in Fig.~\ref{msdstat}b, where the particle translates in the initial $x$ direction before reaching its equilibrium 
configuration.  In contrast to the skimming case, the probability distributions in $y$ turn out to be Gaussian at all times.

\begin{figure} [!htb]
\centering
\includegraphics[width=0.95\textwidth]{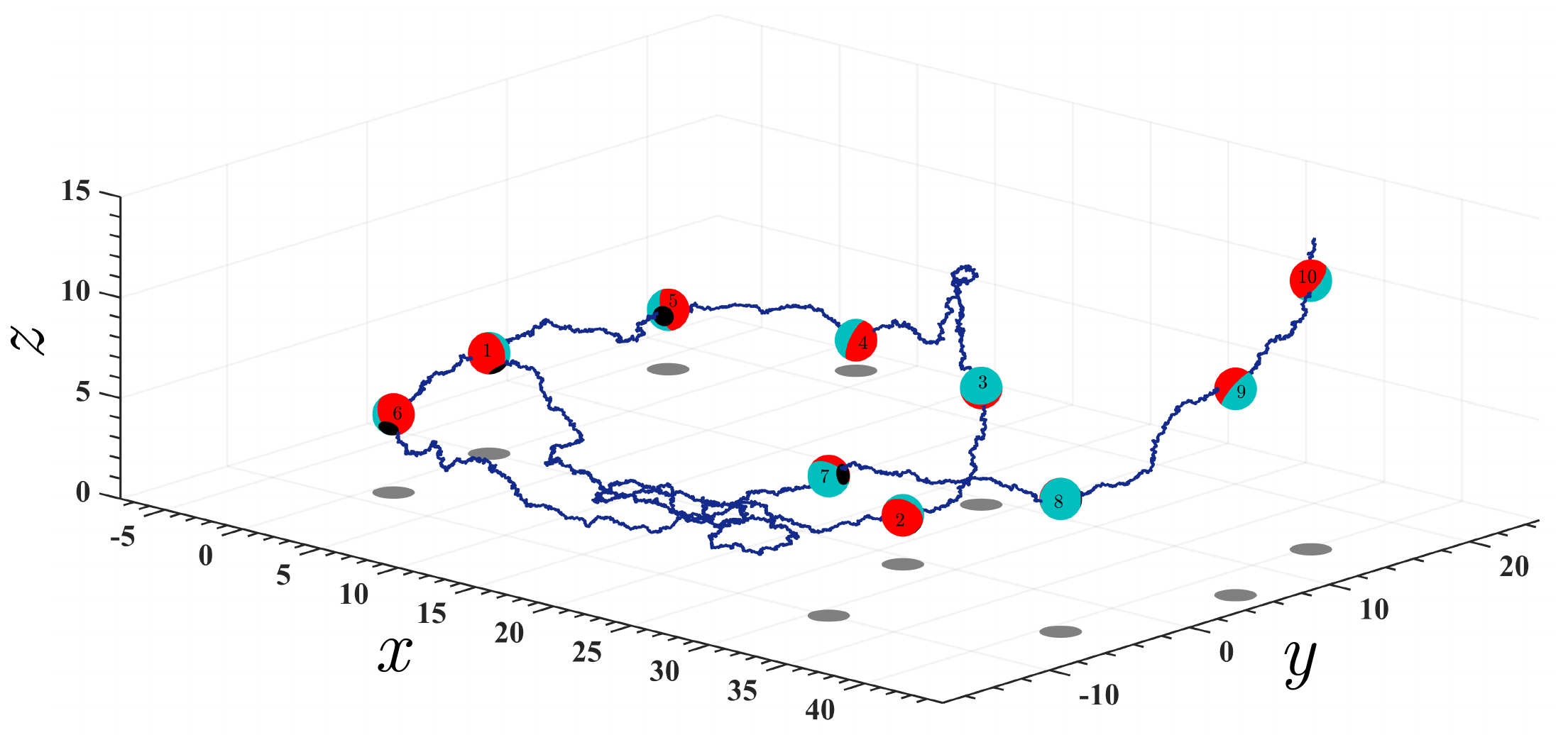}     	    
\caption{\footnotesize{Trajectory of swimmer with $\theta _{cap}  = 150^ \circ$ for small values of P\'eclet number ${\rm P}e=100$.}} 
\label{statp100}
\end{figure}

Decreasing the P\'eclet number enhances the fluctuations, and for values below 200 the particle tends to leave the wall after a transient period, since the (larger) fluctuations in  
orientation eventually bring $\Xi$ below $90^\circ$ and consequently the particle is driven away, until further fluctuations reorient the particle back towards the wall. A sample trajectory at
${\rm P}e=100$ is shown in Fig.~\ref{statp100}, indicating the larger positional fluctuations and the transience of the near-wall state. 

\section {Summary and Conclusion}
This paper has studied the effect of thermal noise on the motion of a diffusiophoretically self-propelled colloid near a no-slip planar wall that does not adsorb solute. We focus on a spherical Janus swimmer with a symmetric catalytic cap as a model engine which autonomously moves in the fluid medium. The mechanism driving the motion of the colloid is neutral self-diffusiophoresis where uncharged solutes with repulsive interactions are produced with constant flux
at active cap of the particle.  Moreover, the thin interaction-layer limit is assumed, which allows us to formulate the mass transfer and hydrodynamics of the problem based on a slip velocity approach.
In this treatment, the gradient of the solute along the surface of the swimmer (arising from the asymmetry of activity at the surface of the particle) creates a force on the fluid in the interaction zone which varies tangentially along the surface and is balanced by fluid pressure outside the interaction layer. This pressure gradient produces a slip velocity tangent to the particle surface towards the catalytic cap and propels the colloid in the opposite direction, at least in the absence of any boundaries. It has been shown previously that in the deterministic limit, a nearby planar wall can drastically change the rectilinear motion of a Janus swimmer, and
for particles with high active coverages skimming and stationary states of motion can arise when the active side faces away from the wall.\par
 
 A linear hydrodynamic stability of skimming and stationary states were examined which showed that these states are stable for small translational and rotational perturbations of the particle.
  
 A particularly relevant perturbation for colloidal particles arises from thermal fluctuations in the liquid, and we have investigated their consequences for skimming and stationary swimmers within a Brownian dynamics framework.  At large values of the P\'eclet number, when the propulsive force dominates thermal forces, the skimming states persist with the particle gliding along the wall with small fluctuations in tilt angle and separation distance
from the wall, with the averages peaked around the deterministic values.  Likewise, at high ${\rm P}e$ the stationary swimmer remains trapped near the wall, with orientation close to normal, where the dominant fluctuations are rotation around the axis normal to the wall and lateral
motion along the wall. The strength of the fluctuations increase as the P\'eclet number falls, so that the skimming and stationary states becomes transient, and at very low P\'eclet the motion becomes unpredictable. \par
  For the skimmer, the transition from ballistic to diffusive motion occurs at times much larger that the rotational diffusion time, where memory of the initial orientation is completely lost, and also exhibits some transient initial non-Gaussian behavior in its positional fluctuations. The stationary swimmer does not show
directed motion and remains in the Gaussian regime at all times, and the particle has a larger diffusivity than a passive particle diffusing near a wall.  The principal conclusion of the paper is thus  that the skimming and stationary states are robust, and can provide a platform for  applications.

\newpage
\appendix 
\section{} \label{append1}
We require the torque on a sphere in the presence of a planar wall, rotating about an axis normal to the wall, with boundary conditions of no-slip on the surface of the sphere 
and the wall. Due to the reversibility of the Stokes equations the force on the particle vanishes and the torque is in the normal direction. This problem was first solved by
Jeffrey \cite{jeffery1915} using the fundamental solution of Laplace's equation in bispherical coordinates, and from this we can express the correction to hydrodynamic torque 
relative to that of particle freely rotating in an infinite medium as
\begin{equation}
f^{rr}_\bot \equiv \frac{T_z}{8\pi\mu a^3\Omega} = \sinh ^2 \beta _0 \sum\limits_{n = 1}^\infty  {n(n + 1)\frac{{{\mathop{\rm e}\nolimits} ^{ - (n + \frac{3}{2})\beta _0 }  - 
{\mathop{\rm e}\nolimits} ^{ - (n - \frac{1}{2})\beta _0 } }}{{\sinh (n + \frac{1}{2})\beta _0 }}}, 
\end{equation}
where $\Omega$ is the angular velocity and $\beta_0 = \cosh^{-1}(\delta+1)$.  This expression was obtained previously by Lee and Leal \cite{lee1980} for a particle rotating near a 
planar fluid-fluid interface.

\section{} \label{append2}
We wish to compare the behavior of the orientational correlation function for diffusing swimmer in two and three dimensions. The direction unity vector $\bf\hat{u}$ satisfies the rotational diffusion equation, and the corresponding Greens' function is 
$$
G_3({\bf\hat{u}},t;{\bf{\hat{u}}}_0,0)=\sum_{l=0}^\infty\sum_{m=-l}^l\,e^{-Dl(l+1)t}Y_{lm}(\theta,\phi)\,Y_{lm}^*(\theta_0,\phi_0),
$$
in three dimensions, where $D$ is the diffusivity and $\theta$ and $\phi$ are the spherical coordinate angles associated with ${\bf\hat{u}}$, and
$$ 
G_2({\bf\hat{u}},t;{\bf{\hat{u}}}_0,0)=\sum_{l=-\infty}^\infty\,e^{-Dlt}\,\frac{1}{{\sqrt {2\pi } }}e^{-il\phi}\,\frac{1}{{\sqrt {2\pi } }}e^{+il\phi},
$$
in two dimensions, where $\phi$ is the polar coordinate angle of ${\bf\hat{u}}$ and $1/\sqrt {2\pi } \exp ( - il\phi )$ is the normalized eigenfunction.  The probability distribution for ${\bf\hat{u}}$ is given
by
in either case by
$$
P({\bf\hat{u}},t)=\int\, d{\bf\hat{u}}_0\,G({\bf\hat{u}},t;{\bf{\hat{u}}}_0,0)\,P_0({\bf\hat{u}}_0),
$$
where $P_0$ is the distribution of initial orientations.  The correlation function of interest is $C(t)=\langle{\bf\hat{u}}(t)\cdot{\bf\hat{u}}(0)\rangle$ 
and the integral over ${\bf\hat{u}}(0)$ selects out the $l=1$ term, which has a time dependance $e^{-2Dt}$ in three dimensions and $e^{-Dt}$ in two. Knowing the propulsion force at the equilibrium separation distance and orientation of colloid (${F_P}({\delta _{eq}},{\Xi _{eq}})$),  the mean square displacement can be obtained as:
\begin{equation}
\left\langle {{{({\bf{r}}(t) - {\bf{r}}(0))}^2}} \right\rangle  = 2t/(3f_\parallel ^{tt}) + 2{(4f_ \bot ^{rr}/(3f_\parallel ^{tt})\,{\rm P}e\,{F_P}({\delta _{eq}},{\Xi _{eq}}))^2}\,({e^{ - t/(8f_ \bot ^{rr})}} - 1 + t/(8f_ \bot ^{rr})). \label{msd2d}
\end{equation}

\newpage


%

\end{document}